\documentclass[manuscript,screen,acmsmall]{acmart} 
\usepackage[acronym,xindy]{glossaries} 
\usepackage[section]{placeins} 
\usepackage[export]{adjustbox} 

\usepackage{tabulary}%
\usepackage{tabularx}%
\usepackage{longtable}%
\usepackage{hhline}%

\usepackage{makecell}%

\usepackage{adjustbox}
\usepackage{multirow}
\usepackage{censor}

\newif\ifcensor

\newrobustcmd{\censorxc}[1]{
\ifcensor 
    \textbf{\{Removed for anonymization\}}
\else
   #1
\fi
}

\renewcommand{\censor}[1]{\censorxc{#1}}

\StopCensoring
\usepackage{dirtytalk}

\newcommand{\xquote}[2]{
\begin{quote}
\textit{\say{#1}} - {#2}
\end{quote} 
}

\makeglossaries

\newglossaryentry{ssp}
{
    name={Social Service Provider},
    text={Social Service Provider},
    plural={Social Service Providers},
    description={An individual focused on providing services targeted at youth to assist them in leading a healthy life}
}

\newacronym[plural=Social Service Providers, firstplural=Social Service Providers (SSPs)]{sspacr}{SSP}{Social Service Provider}

\newglossaryentry{cmc}
{
    name=Centralized and Multi-disciplinary Collaboration,
    description={Individuals and/or organizations from different work disciplines and businesses working together towards a common goal under a central body}
}

\newacronym[plural=CMCs,firstplural=Centralized and Multi-disciplinary Collaborations (CMCs)]{cmcacr}{CMC}{Centralized and Multi-disciplinary Collaboration}

\newglossaryentry{mosafely}
{
    name={MOSafely},
    text={Modus Operandi Safely},
    description={Modus Operandi Safely (MOSafely.org) is an open-source community that leverages evidence-based research, data, and Artificial Intelligence to help youth and young adults engage more safely online}
}

\newacronym{mosafelyacr}{MOSafely}{Modus Operandi Safely}

\newglossaryentry{hcai}
{
    name=Human Centered Artificial Intelligence,
    description={Creating and using Artificial Intelligence in a way which enhances human life in contrast to displacing humans}
}

\newacronym{hcaiacr}{HCAI}{Human Centered Artificial Intelligence}

\newglossaryentry{AI}
{
    name=Artificial Intelligence,
    description={Computer technology which tries to replicate or exceed human levels of cognition and decision making}
}

\newacronym{aiacr}{AI}{Artificial Intelligence}

\newacronym{nsfacr}{NSF}{National Science Foundation}

\newacronym{hciacr}{HCI}{Human Computer Interaction}

\newacronym{mlacr}{ML}{Machine Learning}

\newacronym{misdacr}{MISD}{MOSafely \textit{Is that Sus?} Dashboard}

\newacronym{gdpracr}{GDPR}{General Data Protection Regulation}

\newacronym{ipacr}{IP}{Intellectual Property}

\newglossaryentry{openboundary}
{
    name=Open-Boundary Innovation,
    text={open-boundary},
    description={An organization using input and feedback from external and internal sources from the perspective of the organization to get ideas that form the basis of innovative cycles while retaining ownership and control of the product or service that is created}
}

\newglossaryentry{opensource}
{
    name=Open-Source Innovation,
    text={open-source},
    description={The design, development, and control of a product or service by a collective community of individuals}
}

\newacronym{ossacr}{OSS}{Open-Source Software}

\newacronym{isacr}{IS}{Information System}

\newacronym{acmacr}{ACM}{The Association for Computing Machinery (ACM)}

\glsdisablehyper

\newacronym{ietfacr}{IETF}{Internet Engineering Task Force}

\newacronym{fpaacr}{FPA}{Fair Play Alliance}

\newacronym{cseaacr}{CSEA}{Child Sexual Exploitation and Abuse}

\newacronym{csamacr}{CSAM}{Child Sexual Abuse Material}

\newacronym{csemacr}{CSEM}{Child Sexual Exploitation Material}

\newacronym{rfcacr}{RFC}{Requests for Comments}

\newacronym{ieeesaacr}{IEEESA}{Institute of Electrical and Electronics Engineers Standards Association}

\newacronym{isoacr}{ISO}{International Organization for Standards}

\newacronym{w3cacr}{W3C}{World Wide Web Consortium}

\newacronym{nmecacr}{NCMEC}{National Center for Missing \& Exploited Children}

\newacronym{interpolacr}{INTERPOL}{The International Criminal Police Organization}

\newacronym{iwfacr}{IWF}{Internet Watch Foundation}

\newacronym{wefacr}{WEF}{World Economic Forum}

\newacronym{centracr}{CENTR}{Council of European National Top-level Domain Registries}

\newacronym{igfacr}{IGF}{Internet Governance Forum}

\usepackage{xcolor}
\AtBeginDocument{\colorlet{defaultcolor}{.}}
\AtBeginDocument{\colorlet{editcolor}{black}}

\newbool{editing}

\newrobustcmd{\edit}[2][editcolor]{%
\colorlet{currentcolor}{.}%
\ifbool{editing} {%
    \leavevmode\color{#1}{#2}%
    \color{currentcolor}%
    }%
    {%
        \leavevmode\color{defaultcolor}{#1}%
    }%
}%

\booltrue{editing}

\AtBeginDocument{%
  \providecommand\BibTeX{{%
    \normalfont B\kern-0.5em{\scshape i\kern-0.25em b}\kern-0.8em\TeX}}}

\setcopyright{rightsretained} 
\copyrightyear{2024}
\acmYear{2024}
\acmDOI{XXXXXXX.XXXXXXX} 

\acmConference[CSCW'25]{Computer Supported Cooperative Work and Social Computing}{October 18--22,
  2025}{Bergen, Norway}
\acmISBN{978-1-4503-XXXX-X/18/06} 

\begin{document}

\title{Building a Village: A Multi-stakeholder Approach to Open Innovation and Shared Governance to Promote Youth Online Safety}


\author{Xavier V. Caddle}
\email{xavier.caddle@ucf.edu}
\orcid{0000-0002-8632-2283}
\affiliation{%
  \institution{University of Central Florida}
  \streetaddress{4328 Scorpius St}
  \city{Orlando, FL}
  \country{U.S.A}}

\author{Sarvech Qadir}
\affiliation{%
  \institution{Vanderbilt University}
  \streetaddress{2201 West End Ave}
  \city{Nashville, TN}
  \country{U.S.A}}
\email{sarvech.qadir@vanderbilt.edu}
\orcid{0000-0000-0000-0000}

\author{Charles Hughes}
\email{charles.hughes@ucf.edu}
\orcid{0000-0002-2528-3380}
\affiliation{%
  \institution{University of Central Florida}
  \streetaddress{4328 Scorpius St}
  \city{Orlando, FL}
  \country{U.S.A}}

\author{Elizabeth A. Sweigart}
\affiliation{%
  \institution{Vanderbilt University}
  \streetaddress{2201 West End Ave}
  \city{Nashville, TN}
  \country{U.S.A}}
\email{elizabeth.sweigart@vanderbilt.edu}
\orcid{0000-0003-1287-6774}

\author{Jinkyung Katie Park}
\affiliation{%
  \institution{Clemson University}
  \streetaddress{105 Sikes Hall}
  \city{Clemson, SC}
  \country{U.S.A}}
\email{jinkyup@clemson.edu}
\orcid{0000-0002-0804-832X}

\author{Pamela J. Wisniewski}
\affiliation{%
  \institution{
Socio-Technical Interaction Research Lab}
\city{Orlando, FL}
  \country{U.S.A}}
\email{pamwis@stirlab.org}
\orcid{0000-0002-6223-1029}

\renewcommand{\shortauthors}{Caddle, et al.}
\renewcommand{\shorttitle}{Building a Village}
\begin{abstract}
  The SIGCHI and Social Computing research communities have been at the forefront of online safety efforts for youth, ranging from understanding the serious risks youth face online to developing evidence-based interventions for risk protection. Yet, to bring these efforts to bear, we must partner with practitioners, such as industry stakeholders who know how to bring such technologies to market, and youth service providers who work directly with youth. Therefore, we interviewed 33 stakeholders in the space of youth online safety, including industry professionals (n=12), youth service providers (n=11), and researchers (n=10) to understand where their visions toward working together to protect youth online converged and surfaced tensions, as well as how we might reconcile conflicting viewpoints to move forward as one community with synergistic expertise on how to change the current sociotechnical landscape for youth online safety. Overall, we found that non-partisan leadership is necessary to chart actionable, equitable goals to facilitate collaboration between stakeholders, combat feelings of isolation, and foster trust between the stakeholder groups. Based on these findings, we recommend the use of open-innovation methods with their inherent transparency, federated governance models, and clear but inclusive leadership structures to promote collaboration between youth online safety stakeholders. We propose the creation of an open-innovation organization that unifies the diverse voices in youth online safety to develop open-standards and evidence-based design patterns that centralize otherwise fragmented efforts that have fallen short of the goal of effective technological solutions that keep youth safe online.


\end{abstract}



\received{2 July 2024}
\received[revised]{10 December 2024}
\received[accepted]{12 February 2025}

\maketitle

\section{Introduction}
With 93\% of U.S. teens having access to the internet through smartphones~\cite{Anderson2023}, youth are heavy consumers of internet-connected technologies. This access exposes them to myriad risks, including cyberbullying, sexual risks, and inappropriate material~\cite{Baldry_school_2017, Razi_lets_2020, Badillo_children_2020}. 
Youth online safety has been a central concern to the Human-Computer Interaction (HCI) research community, with studies ranging from digital parenting and family communication, to teen self-regulation (e.g., \cite{rutkowski2021family, akter2022parental, Schoenebeck2021, wisniewski2017parental, davis2023supporting}) as means to safeguard youth online. However, parent- and family-centric approaches to youth online safety often assume a significant level of privilege, as they require considerable parental time and attention and they may be influenced by other aspects of the parent-child relationship dynamic~\cite{wisniewski2022privacy}. Regardless, the voices of these primary stakeholders, youth and their parents, have received the bulk of coverage in academic studies. 
Yet, more recent work has highlighted that the protection of teen privacy and the promotion of teen online safety via sociotechnical systems is only possible when the larger community of policymakers, advocacy organizations, educators, technologists, scholars, parents, and youth work together as part of a broader social-ecological support ecosystem ~\cite{caddle2024}. Indeed, the voices of the secondary stakeholders in youth online safety have not received much attention~\cite{caddle2022, caddle2024}. Both researchers and Internet-related organizations have emphasized the importance of including multi-stakeholder perspectives~\cite{buckridge2024} and making concerted efforts to bring stakeholders together to discuss opportunities and challenges related to internet-related grand challenges~\cite{NETmundial2014,igf2024,NETmundial2024}. 
As such, the extant literature and political landscape around youth online safety highlight an urgent need for a unifying approach that brings together the voices of secondary stakeholders, such as industry professionals, healthcare providers, and academia to foster dialogue and innovation related to youth online safety~\cite{caddle2022, caddle2024}, so that we can move the field and practical application of our efforts forward. Grounded in the computer-supported collaborative work (CSCW) ethos, we looked to open innovation--an approach used by organizations to intentionally leverage both internal and external input to create and deliver products and services~\cite{chesbrough2003}--as a model for bringing together the insights and inputs of secondary youth online safety stakeholders to fill this important gap in the literature. In other sociotechnical contexts, open innovation has been explored as a collaborative way to promote common goals, including social good~\cite{ahn2019leveraging, chesbrough2006open, lukas2024}.  

We interviewed 33 participants from key stakeholder groups to understand what is needed for a sustainable youth online safety community. Leveraging design probes to elicit stakeholder feedback, we examined the type of community needed and the types of outputs desired from these stakeholders. The design probes comprised of two major components: 1) the \censorxc{\acrlong{misdacr} (\acrshort{misdacr})~\cite{alsoubai2022}}, a youth-centered risk detection dashboard, as a potential work product of the proposed community, and 2) a community platform that allows members to view news, posts, and projects related to youth online safety, and to participate in discussions with other community members. 
In our multi-stakeholder interview study leveraging these design probes, we focused on answering the following research questions: 

\begin{itemize}
    \item \textbf{RQ1: \textit{Tensions}} - \textit{What tensions arise when these stakeholders are asked to envision an open innovation community for the purpose of youth online safety?}
    \item \textbf{RQ2: \textit{Reconciliation}} - \textit{How can these tensions be resolved, so that a synergistic multi-stakeholder approach can advance the field of youth online safety?}
\end{itemize}

By answering these questions, we bring into focus the cross-disciplinary viewpoints of secondary stakeholders in youth online safety to identify collaborative opportunities towards building a cohesive open innovation community for addressing the problem of youth online safety. 
This study makes the following contributions to the CSCW and SIGCHI communities: 1) Demonstrates a novel synthesis of open innovation in the context of youth online safety communities,  
2) Highlights the tensions among stakeholder groups in youth online safety, while charting a path forward to overcome these barriers using open innovation with shared governance, and 
3) Provides evidence-based guidance for the development of an inclusive and equitable multidisciplinary youth online safety open innovation community. 

As such, our work fills a critical gap in CSCW literature by demonstrating how open innovation and shared governance can be synthesized within the context of youth online safety communities, contributing to domain-specific CSCW applications and exemplifying boundary-crossing by integrating insights from social computing, open innovation, governance practices, and youth safety. This interdisciplinary approach transcends traditional research silos, fostering a holistic understanding of CSCW systems’ potential to enhance online safety and aligning with the CSCW community's goal of designing, evaluating, and understanding technologies that support social and cooperative practices.
Finally, our research advances the design of computing technologies for cooperative work and social interactions by advocating for open innovation and shared governance. This approach fosters inclusive, equitable collaboration and transparent decision-making, offering new insights into the creation of supportive environments and governance practices critical for effective youth online safety solutions in CSCW not previously explored in the literature.

\section{Background}
First, we build a case for the need for an open innovation community for youth online safety based on the literature on open innovation communities focused on a common goal, particularly to promote social good that benefits society. Then, we synthesize past work that has uncovered key challenges with an open innovation approach to engage with and resolve tensions among key stakeholders. 
\subsection{A Case for Building an Open Innovation Community for Youth Online Safety}
Online interaction has become a vital part of youth culture, making youth online safety a central concern of Human-Computer Interaction (HCI) research, ranging from identifying the serious risks youth face online to developing evidence-based interventions to reduce these risks (e.g., \cite{agha2023strike, razi2023sliding, alsoubai2022friends, akter2022parental, rutkowski2021family, Schoenebeck2021,  masaki2020exploring, davis2023supporting, alluhidan2024teen, park2024personally, park2023towards}). Early work on youth online safety emphasized restrictive mediation like parental control apps limiting access to social media or setting content rules for media and exposure~\cite{schiano2017parental, khurana2015protective}. These approaches assume privilege, requiring significant parental time and attention, and are influenced by parent-child relationship dynamics~\cite{wisniewski2022privacy, wisniewski2025moving}. 
Recent efforts advocate designing socio-technical systems to involve secondary stakeholders in youth online safety~\cite{badillo2024towards, caddle2024, caddle2022, qadir2024towards, sweigart2025takes}. 
For instance, social service providers (SSPs) face challenges relying on youth self-reported data for risk experiences and suggest AI for early detection but worry about digital trace data access and trust issues. The SSPs, however, thought such a solution could not be implemented due to a lack of access to digital trace data and concerns about violating the trust relationships they have built with youth.
~\cite{caddle2022}. Researchers have explored tensions between privacy and protection among stakeholders (e.g., IT professionals, clinicians, educators) and highlighted the need for broader community collaboration~\cite{caddle2024, caddle2022, sweigart2025takes}. Promoting teen online safety requires policymakers, advocacy organizations, educators, parents, scholars, and youth to work together in a broad social discourse~\cite{caddle2024}. Collaborative approaches like open innovation can facilitate this discourse toward promoting social good. Caddle et al.~\cite{caddle2024} pointed to centralized multidisciplinary communities using open algorithms to achieve this in the youth online safety space but did explore the implementation of such a community. We build upon these prior works by investigating what is necessary for such a community to be successful.


Open innovation refers to an organization intentionally using input from both within and outside of itself to create and deliver its products and/or services~\cite{chesbrough2003}. 
For organizations, “working open” offers benefits including increased \textit{spread}, \textit{demand}, \textit{complementarity}, \textit{privileged information}, and \textit{judgment} to enhance organizational profit~\cite{weiss2015}. These benefits have garnered much interest in the academic and secular communities~\cite{lee2009}, as well as in governance~\cite{simon2005, weiss2015}. For example, Weiss et al. demonstrated that companies could better address cybersecurity threats with open-source methodologies compared to traditional siloed approaches~\cite{weiss2015}. 
Prior work ~\cite{eppinger2021open, chesbrough2014open, svirina2016implementing, ahn2019leveraging, chesbrough2006open} shed light on open innovation as a means to promote social good. They highlighted the practices such as open-source co-creation platforms and product development partnerships that could enable innovation that delivers high societal benefits~\cite{eppinger2021open, ahn2019leveraging}. 
By engaging users in the innovation process, scholars showed that open innovation can facilitate knowledge diffusion and shape desired outcomes~\cite{eppinger2021open}. As such, prior work advocated for multi-stakeholders, including companies, innovators, funders, citizens, and politicians, to direct efforts toward socially beneficial innovations~\cite{chesbrough2006open, eppinger2021open, ahn2019leveraging}.
Using an open innovation methodology could be beneficial to the youth digital well-being~\cite{FTA2022combatting, OSG2023_advisory} as it could mitigate the issues of perceived opacity, siloed implementations, duplication of efforts, and the non-inclusion of diverse stakeholder voices. Despite its potential, implementing open innovation for youth online safety remains under-explored.

\subsection{Overcoming Key Challenges of Working Open}
Working open with diverse stakeholders is not without challenges. Prior research showed that open projects could suffer from a lack of engagement, loss of key leadership ~\cite{shibuya2009}, internal cultural differences, external market factors, and interpersonal conflicts often stemming from cultural differences and biases~\cite{rruong2022,davidson2014}, all of which have resulted in low success rates~\cite{schweik2012}. 
Shchweik et al.~\cite{schweik2012} investigated approximately 174,000 open-source projects on SourceForge, a popular software-sharing site, and identified the major reasons for open-source project failure: 1) losing the development team or lack of developer interest, 2) resource constraints (money, time, people), 3) team conflicts, 4) project not being kept up-to-date or lagging behind its competitors, 5) unconstrained project growth, 6) poor or non-existent quality control, and 7) legal problems. 
To overcome some of the challenges of working open, Lee et al.~\cite{lee2009} demonstrated the \acrshort{ossacr} success model focused on features such as software quality, user satisfaction, individual net benefits, and community service quality. Similarly, Ghapanchi et al.~\cite{ghapanchi2011} provided six practical measures to consider when determining the success of \acrshort{ossacr} projects and products. These measures are 1) \textit{user interest}, 2) \textit{project activity}, 3) \textit{project effectiveness}, 4) \textit{project performance}, 5) \textit{project efficiency}, and 6) \textit{product quality}. In pointing to these measures, Ghapanchi et al. encouraged the \acrshort{ossacr} community to consider multiple measures rather than simplistic measures related to the development environment~\cite{ghapanchi2011}.

Understanding secondary stakeholder perspectives and needs regarding collaborative efforts in youth online safety is important to ensure the longevity of collective outcomes and thus long-lived youth online protections. More importantly, it is critical to address different needs and tensions among stakeholders that could undermine open efforts in youth online safety, potentially leaving youth in vulnerable positions online. 
Yet, the challenges and potential solutions to reconcile challenges in working openly toward youth online safety community building have been under-explored in the CSCW and broader SIGCHI communities. To fill this gap in the literature, we engaged with 33 secondary stakeholders in youth online safety and investigated 1) their needs and tensions in collaborative opportunities and 2) solutions to reconcile tensions to achieve a more cohesive approach to foster dialogue and innovation using open methodologies toward promoting youth online safety. 

\edit{\subsection{Coordinating Multistakeholders in Collaborative Work}
Researchers in the CSCW community have highlighted the challenges of conducting collaborative work such as fragmentation of efforts teams leading to duplicative or incompatible solutions~\cite{Nomura2008}, divergent priorities among team members which hinder trust and coordination \cite{freeman2012values, sorensen2009making}, unequal decision-making power among stakeholders leading to feelings of isolation or marginalization \cite{sorensen2009making}, and that differing roles and positions in hierarchical organization structure (functional and organizational distance respectively) are predictive of productivity~\cite{Wang2022}. The CSCW community has used Olson's four-dimensional framework~\cite{Olson2000} to analyze or predict the success of a collaboration~\cite{Ahmed2012, Mao2019}. The framework's four dimensions are 1) nature of work, 2) common ground, 3) collaboration readiness, and 4) technology readiness. Nature of work refers to how easily the work can be divided to be completed by separate teams. Successful distributed collaboration necessitates loose coupling of deliverables worked on by remote teams under Olson's framework. This is not always achievable in research settings where there is much uncertainty~\cite{Nomura2008}. Common ground refers to the team members having a shared knowledge base and vocabulary. This can be complex as collaborative work across domains and disciplines can suffer from the ``dual goal dilemma'' where there is tension within the team to find common ground based on differing expectations and agendas~\cite{Hou2017, Mao2019}. Collaboration readiness refers to the predisposition of the team to collaborate based on skill, and the culture of the team/organization (e.g. competitive vs cooperative). Technology readiness encapsulates the level of technological competence team members have in the collaborative tools being used within their context. Previous research on contemporary multidisciplinary collaborations has found that technology readiness is context-dependent~\cite{Mao2019}. 

Much of the literature on collaboration focused on analyzing particular collaborative tools or (new) features used by multidisciplinary groups or the challenges faced by users of those collaborative tools \cite{feng2023understanding, kashfi2017integrating, kang2024challenges, das2024comes}. Our goal in this study was to move past the tools for collaboration toward mission statements and strategies for multidisciplinary youth online safety collaborative work. Researchers in the youth online safety community are beginning to pay closer attention to the voices of secondary stakeholders, recognizing they have a lesser heard voice in the space while they in turn strive to perform a social good~\cite{caddle2022,caddle2024}.  This research focuses less on increasing the efficiency of current collaborative tools used in youth online safety, but more on understanding the motivations towards collaboration across secondary stakeholders in youth online safety. It imperative to investigate the issues that may lead to distrust among secondary stakeholders in the youth online safety space to improve collaborations~\cite{fisk2016framing}. This study thus focuses more on dimensions 1 (nature of work), 2 (common ground), and 3 (collaboration readiness) of Olson's framework~\cite{Olson2000}.
}

\section{Methods}
We conducted 33 semi-structured interviews using design probes with secondary stakeholders in youth online safety over six months from November 2023 to April 2024 in the United States. The sessions were conducted online using Zoom videoconferencing technology. 

\subsection{Semi-Structured Interview Study with Design Probes}
\begin{figure}[!htbp]
\centering

\includegraphics[scale=.45]{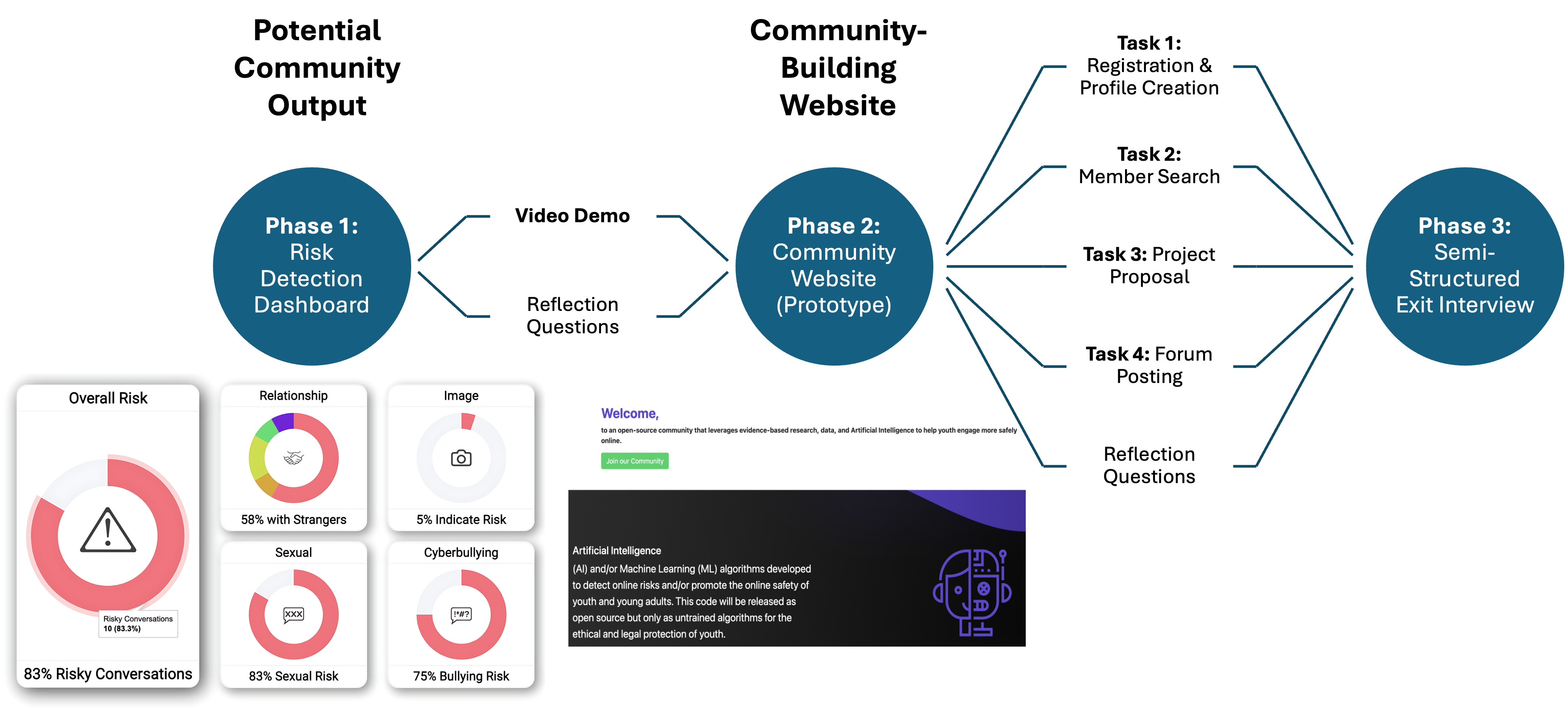}
\caption{Session Flow of Design Probes and Community Tasks}
\Description{The study was comprised of two major sections which included a video demonstration of a risk detection dashboard as a possible community outcome, followed by a series of tasks on an online community created for the study. The tasks included registration and profile editing, searching for members to collaborate with, proposing a project to the community, and posting on a forum.}
\label{fig:study3flow}
\end{figure}

We conducted semi-structured interviews with secondary stakeholders and leveraged two thought-invoking design probes to obtain rich responses from participants. Design probes allow study participants to interact with technology in a more tangible context, while allowing for open-ended feedback to capture insights on the thinking process of participants as they complete tasks~\cite{wallace2013,Shneiderman2016}. Study sessions comprised two main phases: 1) a video demonstration of a youth online risk detection dashboard, and 2) a task completion section asking the participant to complete four tasks via a youth online safety community prototype. In closing, participants were asked their overall views on how to build the community and attract members. The study flow is shown in \hyperref[fig:study3flow]{Figure \ref{fig:study3flow}}. 
This study was approved by the authors' Institutional Review Boards (IRBs). The semi-structured interview questions are included in Appendix \ref{sec.interviewquestions-study3}.

\subsubsection{Youth Risk Detection Dashboard}
\begin{figure}[!htbp]
\centering

{\includegraphics[scale=0.45]{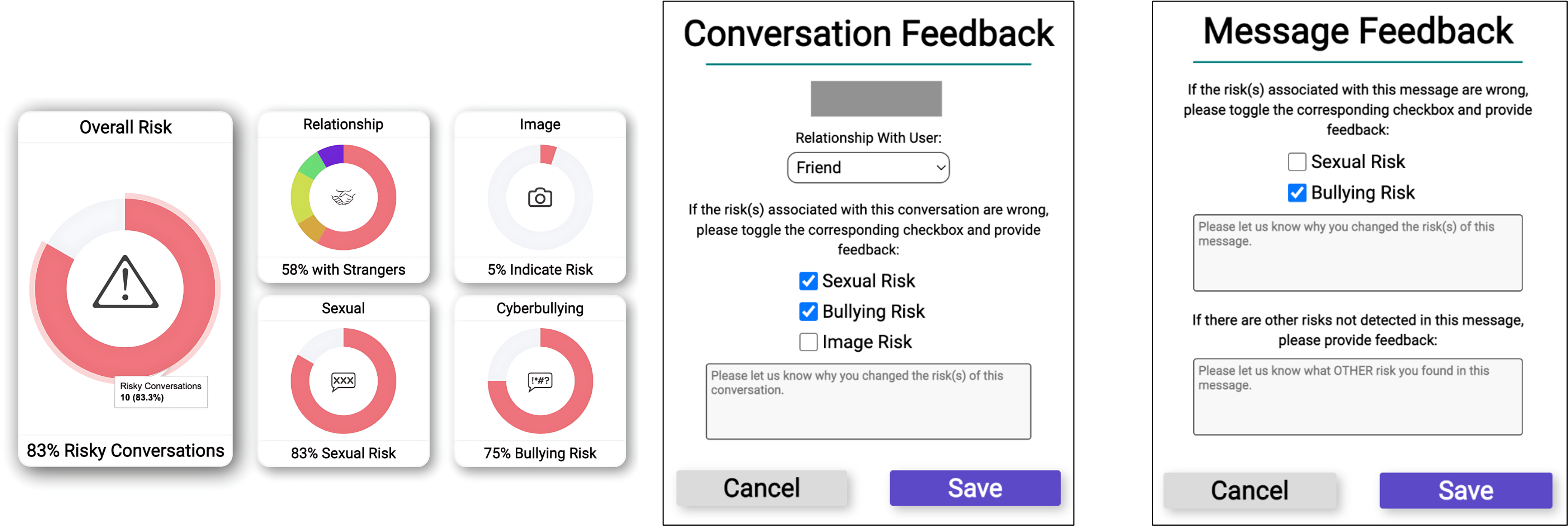}}
\caption{Youth risk detection dashboard showing overall risks and user modification of conversation and message risk flags.}
\Description{This image shows a risk assessment screen with a conversation listing on the left for selection, and the messages within the selected conversation on the right. Youth can select a conversation and view the messages the system identifies as risky.}
\label{fig:study3sec1dash}
\end{figure}

During the dashboard design probe, participants were asked to watch a video demonstration of a youth online safety dashboard~\cite{alsoubai2022}. The purpose of the dashboard in the study was to illustrate an actionable outcome that the community could work on. The dashboard
allows youth to upload social media data files (from Instagram and Twitter) and view \acrshort{aiacr} generated risk assessments based on the contents. The demonstration took the participant through a narrated journey of a teen performing the following steps:

\begin{enumerate}
    \item Dashboard login.
    \item Social media data file upload.
    \item Risk assessment high-level overview showing the amount of risky conversations, and percentage of conversations where cyberbullying and sexual risk are suspected.
    \item Conversation drill-down where the teen can a) change whether the conversation is with a stranger or known associate, b) change whether the overall conversation risk assessment of risky/not-risky is inaccurate, and c) view the messages within the conversation and edit the labels if they perceive the risk message assessment of risky/not-risky is inaccurate (Figure \ref{fig:study3sec1dash}).  
\end{enumerate}

Participants were asked to provide feedback on the concept behind the dashboard after viewing the video demonstration. We probed participants on topics, such as possible usage by other stakeholders, barriers that might prevent community members from contributing to dashboard design, additional features they think might be useful to place on the dashboard, as well as the overall idea of the dashboard as a potential output from contributing community members.

\FloatBarrier

\subsubsection{Community Platform Design}
We created an online community platform as a design probe to interact with youth online safety stakeholders. We added the following features to the community platform based on user needs discovered in prior research~\cite{caddle2022, caddle2024}.  
To support the need for individuals to access scientifically-backed information on youth online safety, we included a news feature. The creation of news items was restricted to administrative users who are vetted experts in the area. To support discussion on youth online safety, we included a posts forum that allows registered users to create discussion topics and comment on them. To support access to youth online safety experts, we included a member search feature allowing stakeholders to find collaborators by using search terms matching name or expertise. Also, we included a project creation feature to allow members to propose projects to the community. Appendix tables \ref{tab:capabilitysupport} and \ref{tab:featureset}  summarizes these modifications.


\paragraph{Community Design Probe Tasks:}
In the community design probe, participants were asked to share their screens to conduct a series of tasks on the online community. These tasks were 1) registering with the community and creating a profile, 2) finding another member they wish to work with based on the profile information of present members, 3) proposing a project to the community, and 4) posting on the community forum. Participants were asked questions about each task before transitioning to the next task.

\begin{itemize}
    \item \textbf{Task 1 -  Registration and Profile Creation:} Participants were asked to use a web browser to navigate the online community to register and create a profile. Participants were notified of successful account creation and asked to log in after successful registration. Upon login, participants were asked to edit their profile where they could choose to edit the information they previously entered (excluding the restricted username data point).
    \item \textbf{Task 2 - Member Search:} The second task was to find and message a community member they would like to work with. The \textit{Users} page of the community site was built to list 10 community member profile cards page and allows users to search for members based on name, area of expertise, or interest. The messaging functionality sent an asynchronous internal (within the community) message to the user chosen by the participant.
    \item \textbf{Task 3 - Project proposal:} Next, participants were asked to create a new project with a project name, a description of the project, and an optional GitHub repository name. After project creation, participants navigated the new project page where they and other community members could comment on the project. 
    \item \textbf{Task 4 - Forum Posting:} The final task study participants were asked to execute was to participate in the forum by creating a post. To create a post, participants were required to provide a post title and post content. After successfully creating the post, participants viewed where they and other community members could interact by commenting directly on the post or to other comments. 
\end{itemize}

\subsubsection{Closing Questions}
At the end of each session, open-ended questions to uncover broader concepts and perspectives on youth online safety, community usage, and possible alternative strategies were asked to complement the community and dashboard design probes. These questions investigated participant knowledge about the existence of alternative technologies that could be used to build the community, what could be added to the community to encourage participation, their perspectives on member-only access to community features, and their views on moderation. These interview questions were chosen to obtain the participants’ views on what needs to be added, modified, and/or removed so as to attract, engage, and retain community members.

\subsection{Study Procedure and Participant Recruitment}
For participant recruitment, we targeted individuals who were 18 years of age or older, spoke English, had previously or were currently working with youth between 13 to 17 years old in a capcity that would inform their online safety. We used non-probabilistic purposive sampling to recruit participants~\cite{Lazar_2017, Tongco_2007}. This sampling method is used for study recruitment during which researchers select participants due to the participant's knowledge of the topic~\cite{Tongco_2007}. We employed the following approaches to reach our target participant count: 1) advertisements placed on social media via the researcher's social media accounts on Facebook, Instagram, LinkedIn, and X (Twitter) and 2) word-of-mouth through the researchers' online and offline social networks. 
We recruited participants from the following groups:

\begin{itemize}
    \item \textbf{Industry Professionals (n=12):} Individuals in entrepreneurial roles or employed in companies focused on technology creation i.e. technology consultants, entrepreneurs, CEOs.
    \item \textbf{Service Providers (SP) (n=11):} Licensed mental health practitioners, psychologists, pediatricians, and educators who work with youth or young adults in the area of online safety or focus on providing training for individuals regarding using the internet safely and transitioning to adulthood.
    \item \textbf{Researchers (n=10):} Conduct research on socio-technical or academic topics related to online safety. All were connected to US universities in research capacities.
\end{itemize}

Prior to participation, an explanation of the research was read to participants, along with the option to withdraw from the study, prior to participation. We acquired informed consent to participate in our study, as well as consent for audio and video recording from all participants before the beginning of the interview sessions. Each interview lasted about 60 minutes. Two interviewers attended the interview with one serving as the lead interviewer and the other taking notes. Participants were thanked for their time and offered a \$20 Amazon gift for their time. Twelve (36\%) of the participants accepted the gift card, while the rest declined. Participant characteristics are included in Appendix \ref{sec.participants-study3}.

\subsection{Data Analysis Approach}
The Zoom interviews were audio and video recorded with the live transcription feature turned on. The first author watched each video and corrected any transcription errors. Our qualitative analysis followed the thematic analysis process described by Braun et al.~\cite{Braun_2019}. The first author having familiarity from conducting and re-watching the interviews, identified the initial dimensions for which to code, and created an initial codebook against the research questions under the direction of senior researchers on the study team. The interviews were then evenly divided between three of the co-authors to code, allowing additional codes to be identified with continuous discussion with the study team. All codes were discussed among the study team, clarified, merged, and organized into themes with differences being resolved through discussion with the senior researchers and consensus building. 
Tables \ref{tab:codebook_rq2-study3-new} and \ref{tab:codebook_rq3-study3} show the themes and sub-themes for the over-arching research questions of this study. Illustrative quotes from the participant responses are used in our results to describe the findings. Each quote ends with the Participant ID (P=Participant) followed by their stakeholder type (e.g., P\#, Researcher).
\section{Results}
In this section, we present the perspectives shared by participants, 
where they had tensions with the concept of youth online safety communities (RQ1). Next, we present what participants viewed as possible resolutions to those tensions to make such communities valuable contributions (RQ2). 

\subsection{Surfacing Tensions: Individual Interests Must Be Protected to Facilitate Collaboration. (RQ1)}
The first research question (RQ1) investigated the tensions that arise when various stakeholders—industry professionals, researchers, and service providers—are asked to envision an open innovation community dedicated to youth online safety. 
Our analysis identified three key themes: 1) differing end goals and 2) varied motivations to contribute to community building, which is shown in {Table \ref{tab:codebook_rq2-study3-new}}.

\begin{table}[htp]\small
\caption{Key Tensions in Stakeholder Goals and Contributions for Youth Online Safety Innovation (RQ1)}
\label{tab:codebook_rq2-study3-new}
\centering

\begin{tabular}{>{\raggedright}p{0.15\textwidth}>{\raggedright}p{0.2\textwidth}p{0.5\textwidth}}
\toprule
\textbf{Theme}                                                      & \textbf{Sub-Themes} & \textbf{Quote} \\ \midrule

\multirow{3}{=}{Stakeholders envision different end goals} 
& \textit{Industry participants want viable products that have sustainable business models} &  \textit{``This is not necessarily a tool that would ever be sort of put into any kind of a market to be used correct? ...So I don't see that there being a product market fit for this in this incarnation in this way that you presented it to me.''} - P25, Industry (Technology Consultant)\\\cline{2-3}
& \textit{Researchers want evidence-based solutions although not launch-ready} &  \textit{``Just, you know, producing knowledge about what teens are encountering and what kind of risky content they're encountering. And then, you know, I could see it being useful for, in the same vein for parents, educators, and technology and yeah, technology companies and policymakers, all of those key stakeholders to get a better understanding of  what teens are encountering online.''} - P15, Researcher \\\cline{2-3}
& \textit{Service providers emphasize awareness and education among parents/youth} &  \textit{``What would catch my attention is something that involves the parents. Okay. that would get my interest, because I feel like the front line, for all of this is for parents to be more aware of the things going on online, what their kids see online, and what their kids have access to online.
''} - P16, Service Provider (Educator) \\ \cline{1-3}

\multirow{3}{=}{Stakeholders have different motivations to contribute to community building} 
& \textit{Industry participants need clear advantages as they already have their own networks} &  \textit{``I guess that I think it all, it just keeps coming down to purpose. What’s the purpose? And why would I be a part of this community as opposed to just being on a Yahoo group or creating my own WhatsApp group with people?''} - P8, Industry (Technology Consultant) \\\cline{2-3}
& \textit{Researchers want connections for impactful research that promotes social good} &  \textit{``I'm thinking, like research interests, if researchers wanted to connect with each other or like even industry, part like, if there's researchers in industry that want to do research?''} - P21, Researcher \\\cline{2-3}
& \textit{Service providers want to connect to provide hands-on support for youth and parents} & \textit{``I am more interested in connecting again, like we out with other parents who have iphone issues security with their kids and asking questions about that.''} - P24, Service Provider (Educator)\\\bottomrule
\end{tabular}
\end{table}

\subsubsection{Stakeholders Envision Different End Goals}\label{study3diffendgoals}

One of the key tensions we identified was the differing end goals envisioned by the various stakeholders involved in the community for youth online safety. These end goals significantly influenced how each group approached the development and implementation of potential solutions.

\textbf{Industry participants aimed to develop viable products with sustainable business models.} When considering the end goals, stakeholders working in the industry were predominantly focused on creating end-to-end solutions that can be brought to market as complete, functional products. Their primary measure of success was the development of products that could generate revenue and have a clear path to sustainability in the market. They exhibited some interplay between altruism and business decision-making. Altruistically, they wanted to do what they can to protect youth online from a technological or business (non-tech) perspective while not incurring a financial loss. At the very least, they desired their expenses to match their income (break-even). Some entrepreneurs shared that they have either pivoted their youth online protective businesses many times due to poor market fit or poor performance, or they completely closed the business and opened another on a different tangent but still within youth online safety. Therefore, community outcomes that did not seem commercially viable were perceived as ``non-starters'' by industry participants as exemplified in the following quote. 
\xquote{We talked with kids, a bunch of kids many of whom were walking around with devices many of whom, you know, engaged and experienced a level of quality and sophistication among the applications they were using on those devices and at home ... things that looked cheap or thrown together, or haphazard, or sort of poorly represented you know, might turn the kids off...}{P25, Industry (Application Developer)}

As a result, the industry professionals were skeptical and critical of the risk detection dashboard we presented as a potential outcome for the community. They did not see it as ``market-ready'' or something end consumers would buy.

\textbf{Researchers wanted evidence-based solutions, being launch-ready was not required.} On the other hand, researchers prioritized evidence-based solutions, even if these solutions were not immediately launch-ready or standalone. They were more open to partial solutions that could contribute to the body of evidence on addressing youth online safety. Researchers saw value in incremental progress and were interested in solutions that, while not fully developed, could provide empirical data to inform future research and interventions. While they were more lenient than their industry counterparts in this regard, researchers emphasized the importance of how these solutions would be evaluated, as their primary concern was the validity and reliability of the evidence produced. If these conditions were met, researchers saw the possible value of contributions to the extraction of ground-truth insights. Overall, researchers viewed partial solutions (such as the risk detection dashboard) as opportunities to gain insights, raise awareness, and develop more comprehensive approaches over time. By focusing on empirical data, researchers aimed to create solutions that were not only theoretically sound but also practically useful in real-world settings. 

\xquote{Certainly help any type of work [from platform] would be something [useful]. The people at Instagram have all this data already. They can. They can do all kinds of things with it. What I think you're trying to do is say, like, we don't know that the data science people at Instagram actually care enough to be sync asking the specific questions we're asking. But the findings can still benefit Instagram.}{P21, Researcher}

\textbf{Service Providers emphasized awareness and education for youth and parents.} Finally, social service providers emphasized awareness and education among parents and youth, with a strong focus on transparency, and little focus on the technology itself. For service providers, the primary goal was to find solutions that could educate and empower both parents and children. In this sense, they evaluated the dashboard design probe as a potential tool for the education of youth, parents, and teachers, as well as for the empowerment of youth. Unlike industry stakeholders, service providers were less concerned with the viability or commercial potential of the solution and more focused on its educational impact. They saw the main value in helping to inform and support the community, regardless of whether the solution can be turned into a marketable product. As such, service providers prioritized solutions that could directly benefit parents and youth by increasing their awareness and understanding of online safety issues, thereby empowering them to make safer choices.
\xquote{This is an opportunity for you [the youth] to take control of your [their] life. You don't have to be in the hands of someone else. So I think the more that we cultivate our belief that they know how to make decisions because telling them not to do it drives them to do whatever it is we don't want them to do.
}{P26, Service Provider (Educator)}

\edit{Divergent end goals created complex dynamics in the community in that }industry stakeholders were driven by the creation of market-ready products, researchers were focused on generating evidence through incremental solutions, \edit{and }service providers prioritized educational outcomes and community support. 

\subsubsection{Stakeholders Have Different Motivations to Contribute to Community Building}

The second key tension we identified was how different stakeholders weighed their contributions to the open innovation community for youth online safety. Below, we unpack how these differences reflected their varied priorities and values, influencing their willingness and manner of participation. 

\textbf{Industry participants want clear advantages to join the community since they already had their own networks.} Industry participants predominantly desired credit for intellectual property (IP) and were keen to contribute to projects where they can maintain a competitive edge. The industry participants' emphasis on protecting IP is rooted in the commercial value and strategic advantage that is provided through proprietary knowledge and innovations. Industry participants shared that during community participation, they probably would intentionally not share information about upcoming products or features (trade secrets) since competitors might also be present in communities. They were amenable to contributing knowledge based on their experience and expertise as long as they were not being asked to divulge trade secrets or information about upcoming product features from their place of employment. 
\xquote{I know there are certain things that I do by day that I can't talk about at night, cause I can't have it linking to competitors ... we want to be first in some of these things.}{P9, Industry (Technology Consultant)} 

Meanwhile, participants from the industry tended to already have established networks and often exhibited a level of distrust towards new or external communities. When discussing joining and contributing to an open community, they expressed a lack of motivation/incentive to do so because they already had networks to rely on. Rather, they were interested in maintaining their existing networks on other platforms and evaluated new community engagements critically to seek clear and purpose-driven benefits such as profit-generating product development (discussed \hyperref[study3diffendgoals]{in earlier section}). This stance ensured that any community participation directly benefits their objectives. For example, industry participants shared that they could not find the benefit of joining the suggested online community (design probe) \textit{without set goals} because they already have an active and well-filtered community in other mainstream platforms. That is, they weighed the purpose and potential benefits of joining a new community against the value of pre-existing communities on contemporary platforms (Yahoo groups, WhatsApp groups, Facebook, etc.). This selective engagement highlighted their preference for communities that provide clear, strategic advantages.

As such, stakeholders from the industry took a cautious approach to sharing ideas and innovations, underscoring their concern about maintaining proprietary advantages. The need to balance collaboration with the protection of unique intellectual property created significant tension within the industry community. This protective stance is evident in their careful consideration of how much to share. In addition, they already had established networks on their own, hence, they needed clear and tangible advantages to contribute to community-building efforts.

\textbf{Researchers wanted connections for impactful research outcomes that promoted social good.}
Participants from the researcher group showed a strong motivation to connect with a diverse range of communities, including those in the industry. They believed that building connections in an open community could create collaborative opportunities, enhance the quality of research, and bridge gaps between academia and industry. 
For instance, researchers envisioned collaborative opportunities across institutions and research streams to find solutions to difficult problems based on contributions submitted to the community. 
They expected that collaboration beyond their field of expertise would significantly enhance the quality of research given the interdisciplinary nature of youth online safety research. In essence, the desire for connectivity for researchers was rooted in the belief that diverse interactions can lead to richer and more impactful research outcomes.

\xquote{Faculty share research and say, hey, you might want to read my paper and this is the important point of the paper also collaboration. If anybody's working in this area, maybe we can collaborate on that sort of stuff that's all I think I would use the system. I'm not a social media person but anything that helps improve research will be good.}{P17, Researcher}


Another strong motivation for the researchers to join the community-building effort was the access to valuable but hard-to-reach ground truth data (often proprietary) to design evidence-based interventions to promote youth online safety.  One of the struggles that researchers shared was the inability to access rich and ecologically valid youth digital trace data which serves as foundations of sociotechnical intervention. 
As their goals are closely tied to producing insights from ground truth datasets, they welcomed the idea of collaborative efforts to collect and use youth-donated social media data as depicted in our risk detection dashboard design probe.  
At the same time, they were concerned about potential data usage without openly communicating about how the data is collected and used to those contributing their data. Therefore, researchers wanted to ensure that community building is aligned with strong ethical standards while pursuing social good. 
Their strong desire for ethical practices in part stemmed from the nature of academia where research ideas are considered intellectual properties. Similar to industry stakeholders, researchers were concerned about having their research ideas stolen and implemented without credit being given for their research. In this sense, they pointed to the importance of transparency and ethical conduct of the community to achieve the common goals (i.e. promoting youth online safety) in a competitive space. Overall, researchers valued open-source approaches and collaborative efforts that prioritize the public good. Their desire for connectivity for researchers was rooted in the belief that diverse interactions can lead to more impactful research outcomes. Yet, they were keen on ensuring that their contributions lead to tangible social benefits and maintain high standards of transparency and ethical integrity.

\textbf{Service providers want to connect to better provide hands-on support to youth and parents.} Service providers desired to contribute their field expertise to directly benefit the community. They prioritized practical connections, focusing on addressing specific issues and offering direct support. For instance, by engaging with parents within the community, service providers perceived that they could share actionable items with parents to keep their teens safe. This hands-on approach stemmed from their day-to-day active involvement in the life of youth as caretakers for varying periods and not wanting them to experience harm during online interactions. 
For instance, caseworkers wanted to share resources (public but possibly lesser-known state or organization-provided resources), especially with youth who have experienced harm, while therapists wanted to share information to enhance youth mental health and possible parenting strategies to improve parent-teen communication on online harms. 
As such, regardless of their specific sense of duty, participants from the service providers group shared a strong commitment to serving the community, particularly, youth and parents.

\xquote{I do have practical experience. I do have application experience. So I think making it so [the community doesn't] prioritize people that are with these you know well-known research institutions, but are able to really prioritize people that have a sense of urgency and who are willing to truly collaborate...}{P23, Service Provider (Case Worker)}

Another clear motivation for service providers to contribute to community-building was to cultivate a sense of ownership and belonging within the community. This desire was shown in their feeling of \textit{isolation} from other stakeholder groups and their willingness to combat isolation by connecting with them. For instance, caseworkers shared that they did not want their voice to be drowned out or held in less regard than other stakeholders who may be more credentialed (e.g., academic degrees) or carry prestigious titles:  

\xquote{A big motivation for myself and for other people I know in this field, being able to connect with ... sometimes when you're doing this work, you can feel really isolating and you can feel kind of in the weeds. So having that support or connection to other people doing this work it for me. It gives me a lot of hope.}{P6, Service Provider (Case Worker)}

\edit{S}ervice provider\edit{s }focused on connecting with other service providers and parents, aiming to offer real-world help and support. In addition, by connecting with the community, service providers envisioned themselves feeling less isolated and making a positive impact in promoting the digital well-being of youth. 
\subsection{Reconciling Tensions: With Strong Leadership, We Will Contribute. But Make It Easier for Us. (RQ2)}

In considering how the identified tensions could be resolved, thereby facilitating a synergistic multi-stakeholder approach to advance the field, three main themes emerged from the data: 1) community-building efforts need to be lightweight and actionable, yet meaningful,  2) diverse contributions should be equally valued, and 3) the importance of non-partisan leadership to set and act upon common goals. These themes are shown in {Table \ref{tab:codebook_rq3-study3}}.

\begin{table}[htp]\small
\caption{Community outcomes need to be actionable and stakeholder voices respected (RQ2)}
\label{tab:codebook_rq3-study3}
\centering

\begin{tabular}{>{\raggedright}p{0.2\textwidth}>{\raggedright}p{0.27\textwidth}p{0.4\textwidth}}
\toprule
\textbf{Theme}                                                      & \textbf{Sub-Themes} & \textbf{Quote} \\ \midrule

\multirow{2}{=}{Community building efforts need to be strategic, actionable, and lightweight} 
& \textit{Provide valuable information that can be acted upon} &  \textit{``If there's a resource center where they can download guides and useful tools and stuff like that. I think there is definitely a very, that's a useful aspect.''} - P31, Industry (Business Owner) \\\cline{2-3}
& \textit{Balancing between level of effort and value of participating}&  \textit{``I mean, I like the fact that right now it's very lightweight. So that's nice, because I've worked on projects where you're asked to put a lot of detail, and then you just kind of quickly shut down like you know what? I'm not even gonna bother''} - P15, Researcher\\\cline{2-3}
& \textit{Partnerships need to be strategic and action-oriented} &  \textit{``I mean a lot of, part of it is reaching out, right? Finding the right people to talk to that are relevant and have their heart in the right place, aren't trying to rip you off.}- P28, Industry (Entrepreneur)\\ \cline{1-3}

\multirow{2}{=}{Diverse contributions should be equally valued to promote collaboration} 
& \textit{Synergistic contributions greater than the sum of their parts}&  \textit{``And so like, there's this thing I know, and things I don't know... But the things I don't know about are things that I think you know [the community] could help with... and that's where the community can come in.} -P30, Service Provider (Educator)\\\cline{2-3}
& \textit{All contributors' efforts need to be recognized to motivate continued participation}&  \textit{``But you know this tells me it's really collaborative and really thoughtful about pulling in, you know, minds and making it fair to everybody.''} - P1, Researcher\\
\cline{1-3}

\multirow{2}{=}{Non-partisan leadership to set common goals is needed} 
& \textit{Community needs common goals to grow organically}&  \textit{``It [youth online safety] is a huge problem, but it's not a huge problem that's going to fix itself within this, within the problem. It's going to have to be attached to the problem, but it can't replicate the problem.''} - P30, Service Provider (Educator) \\\cline{2-3}
& \textit{Non-partisan leadership is necessary to define common goals}&  \textit{``If somebody contacted me and said, ‘Hey, this is what we’re doing, and we need your perspective,’ then I’ll, you know, I’m gonna create as much time as I can''} - P24, Service Provider (Teacher) \\\bottomrule
\end{tabular}
\end{table}

\subsubsection{Community-Building Efforts to be Strategic, Action-oriented, and Lightweight}

Across the stakeholder groups, there was consensus that efforts to build a youth safety community must be strategic, action-oriented, and lightweight. Participants acknowledged numerous constraints on their time, but indicated that they would engage with content that was targeted to their interests and meaningful. 

\textbf{Provide valuable information that can be acted upon.} Across all stakeholder groups, participants emphasized the need for community-building efforts to be both targeted and actionable. 
Although the idea of community underpinning the efforts to keep youth safe online inspired feelings of altruism among participants, their responses suggested a quid pro quo orientation where the immediate benefits experienced by teens using the tool would determine their perception of its value and influence their continued engagement with it. That is, participants tied the practicality of the efforts to their perceived value of them.

\xquote{If it's not really actionable and providing me real value pretty quickly, I don't know why people would continue to use it.}{P11, Industry (Entrepreneur)}

For instance, participants emphasized that providing data for its own sake was not as helpful in giving stakeholder groups actionable insights compared with curating a knowledge base and providing information in digest form. Time constraints were a frequently mentioned limitation across the stakeholder groups. Therefore, participants often elaborated on the need to be provided with well-curated information that can be acted upon. 
In this sense, participants wanted a simple way to determine at a glance what information is relevant to them and how they could add value by engaging with it. For instance, participants suggested periodic updates and email summaries to keep them informed of ongoing activities efficiently. They emphasized the convenience of quickly scanning these updates to determine their relevance. Members of the researcher and provider stakeholder groups were more open to receiving content over email than using an existing social media platform (e.g., Slack, Discord) or a bespoke tool.
\xquote{Kind of like an email summary… being able to kind of scan that and eyeball to see if it's worth my time to go deeper into it. That might be useful here.}{P1, Researcher}


At the same time, participants noted that creating another social media platform without curation or moderation was not the solution. Participants also expressed frustration with existing communication platforms where the volume of messages and postings made it difficult to keep up. This concern was particularly palpable for researchers who already feel inundated with more new literature than they can read or assimilate. As a result, participants voiced that they either avoided engaging with new communities of practice or that they only gave a cursory look at information shared on these platforms. Therefore, due to time constraints and the need for efficiency, participants viewed a centralized moderator and curator as necessary for making the experience worthwhile.



\textbf{Balancing between the level of effort and value of participating.}  Next, participants were seeking opportunities to engage in ways that are straightforward and convenient, ensuring that their contributions delivered maximum value to the community in the limited time they had available. While many participants expressed a strong desire to make meaningful contributions, many of them lamented their lack of time to be more engaged. 

\xquote{I feel like the project is very big. And so it could mean a lot of things. Then maybe that means we need a category here to categorize the different types of projects, right? So that people can find opportunities easier depending on what types of collaborators they're looking for. I don't want people wasting their time reading my project. If they're not who I want to collaborate with}{P21, Researcher}

For many participants, purpose and meaning were intertwined. \edit{P}articipants felt a sense of purpose in the work of youth online safety and strongly believed in its importance. This clear purpose makes their efforts feel meaningful and motivates them to continue, even when faced with challenges with limited time to contribute. \edit{T}hey wanted to ensure that contributing to community-building would deliver benefits above and beyond what they would achieve on their own. 
In this sense, participants expressed a need for more integrated solutions to interact with community members. They acknowledge that creating a new platform rehashing features commonly associated with current social media platforms could potentially overwhelm community members. The need for integration was further reinforced by participants stating that they generally now spend little time using social platforms even if they were previously avid users.
\xquote{I would just choose probably Discord or Slack, or one that needs email integration cause a lot of folks just won't go.}{P33, Industry (Entrepreneur)}
\edit{P}articipants indicated a desire to strike a balance between effort and value in determining their participation in the community \edit{denoting }a desire for more straightforward, lightweight, and convenient ways to engage that maximize their limited time. 

\textbf{Partnerships need to be strategic and action-oriented.} Another trend we observed was the need for building strategic and action-oriented partnerships. Our participants expressed that, to form strategic partnerships, they need to have standards for collaboration with trusted peers and established experts. The idea of having standards for both the information contributed as well as vetting community members was echoed by numerous participants. Suggestions offered by participants included establishing guidelines for membership, verification of credentials, and evaluation of the quality of contributions as elaborated in the following quote:
\xquote{This is supposed to be a community of people working together. Well there should be some guidelines … protect our community. Yeah? There's either 1 or 2 ways to go, right? If you either kick people out by their standards, and what they produce, or you vet them before they go on.}{P5, Industry (Entrepreneur)}

Participants from the industry stakeholder group were particularly concerned with the reputation of prospective community members and, hence wanted to be able to validate members' qualifications and associations for individuals not already known to them. For participants from the industry cohort, there was an assumption of quality with respect to researchers and community members from academic institutions, but wariness about others with whom they did not have a preexisting relationship or reputational awareness.
Therefore, 
finding a strategic balance between mission-oriented youth online safety work and commercial initiatives was considered important to industry participants. 
For researchers, the youth online safety community could be a place to find potential collaborators not only for specific research projects or grant proposals but also for further community-building efforts through conferences to strategically improve the research outcomes. Therefore, they expressed a need to build a well-structured online community to share what collaborative efforts are needed for which tasks with community members.

\xquote{I guess if they're on this site [open community], then that helps me narrow down. like, at least I know that they're interested in this kind of space like the teams and online safety. And so if I'm looking for a collaborator, or maybe I'm writing a Grant and I need people on my advisory board or if I'm organizing a panel at a conference I might reach out here."}{P 15, Researcher}

In this sense, participants gravitated toward having open forums for information sharing and drawing from diverse experts in their respective fields. They wanted these spaces to be strategically governed, but not censored, advocating for including credentialed individuals who will share accurate and evidence-informed material. 
Overall, the participants emphasized the significance of forming connections with relevant and trustworthy individuals and stressed the necessity of ensuring that partnerships within the community are both strategic and actionable. They highlighted that connections within the community should not only facilitate professional or academic growth but also be easy to maintain and genuinely beneficial for all parties involved.

\subsubsection{Diverse Contributions Should be Valued Equally to Promote Collaboration}
The second major theme we identified was diverse contributions from different stakeholders should be equally valued and rewarded to motivate participation. \edit{Below, these sub-themes are unpacked}. 

\textbf{Synergistic contributions should be greater than the sum of their parts.} 
Stakeholders broadly acknowledged that the whole could be greater than the sum of its parts as, despite the parts being small efforts by some, in concert with others, significant benefits could result.
Participants perceived that diverse contributions \edit{from } different stakeholders \edit{would create } cohesive and robust outcomes, amplifying the overall impact of the collective effort. When discussing the importance of synergistic contributions from diverse members, participants often evoked the principles of open innovation, especially the collaborative efforts that leverage the unique strengths and perspectives of various stakeholders. 
For instance, industry professionals can bring practical insights into technological capabilities and constraints, enabling the development of realistic solutions. Service providers, who often interact directly with youths, could offer valuable on-the-ground perspectives that can inform user-focused safety features. Researchers can contribute empirical evidence and theoretical frameworks that help understand the risks and effective interventions. By valuing these varied contributions, participants envisioned that an open innovation approach can facilitate the creation of comprehensive, robust, and adaptive strategies that enhance online safety in ways that are informed by technology, grounded in real-world application, and supported by scientific research.
\xquote{You end up with questions that would be of interest to a group. And, to put it out there. Maybe it's a measurement question. So it's specific to researchers. But there's also potentially questions that would be valuable to be able to ask other professional groups.}{P7, Researcher}

\edit{A}lthough some questions or issues may be particularly relevant to one subset of the community (e.g., researchers), participants perceived there is also potential for these questions to be of interest to and, hence, answered by other professional groups. This opens up the possibility for cross-disciplinary collaboration, where diverse professional perspectives can contribute to, be appreciated, and enrich the further questions being explored.

\textbf{Recognize all contributors' efforts through open communication to motivate continued participation.} Regardless of the form of contribution, participants noted that it is important for the community to recognize and value all contributions to keep stakeholders engaged and incentivized. Participants frequently showed a strong appreciation for the open innovation principles of transparency and open communication to motivate continuous collaboration among diverse stakeholder groups. 
For instance, some participants shared that there is a misconception about the open-source project that could be resolved through transparency and open communication. One participant noted that misunderstandings about open-source software often label it as commercially nonviable, assuming it cannot generate revenue. This disconnect could result in industry stakeholders choosing not to participate in the community because they believe that they will not be rewarded for their contributions. Therefore, feedback from various stakeholder groups consistently highlighted the importance of clear communication not only regarding management and governance of contribution but also potential compensation and commercialization opportunities followed by their contribution. Participants recognized that ensuring the community is well-informed on these matters is essential for fostering engagement and trust.
\xquote{There's a lot of confusion between open-source and free. There's the perspective that engaging in an open-source project means giving away one's intellectual property and therefore losing rights to commercialize or monetize their work.}{P14, Researcher}

\edit{P}articipants showed that the success of community-driven initiatives relies on the engagement and retention of diverse stakeholders, each bringing unique motivations and backgrounds. 
Recognizing and equally valuing these different contributions is crucial for fostering a sense of inclusivity and respect within the community. When each member feels that their specific interests and goals are acknowledged and appreciated, they are more likely to remain actively involved and committed to the community's collective objectives. This not only enhances the project's richness and diversity but also ensures a robust, multi-dimensional approach to problem-solving, where commercial success and innovative research mutually benefit from each other's strengths.

\subsubsection{Non-partisan Leadership to Collaborate Toward Common Goals is Necessary:} The final major theme captured stakeholders' overarching views on how the community should work toward common goals. The two sub-themes we identified under this theme were: 1) common goals are needed to grow organically and 2) non-partisan leadership is necessary to define common goals.

\textbf{Community has to grow organically with common goals.} 
All participants viewed youth online safety as an important topic and perceived efforts in this stream to be worthy of commendation. Participants thought the design of youth safety technology being informed by inputs from broader society was important. The underlying idea shared by participants was that \textit{something} is needed to help improve the complex relationships that exist between youth, parents/guardians, and technology, given the perceived harm youth are exposed to online. They thought uncovering what must be done requires a community viewpoint because of the prevalence of youth internet usage and varying views of the public. Regardless of the primary interest of any stakeholder group, they viewed the other groups as necessary in the creation and promotion of safety efforts to enhance end-user trust.
\xquote{I don't think that there's a person walking the street that does not have an opinion about device driven life now.}{P26, Service provider (Educator)}
While recognizing the overarching common goals, participants expressed apprehension about randomly reaching out to others, even if they knew they were working in the same field. Instead, they preferred reaching out to, and being reached out to, when there was some commonality between the parties involved. As such, they discussed the community being grown \textit{organically}, that is, through working on common goals. They did not see themselves, or others, actively searching for projects in the space of youth online safety to work on. In discussing this, they noted feeling isolated in their efforts to protect youth as they execute the duties of their stakeholder role. Youth online safety communities providing opportunities to network \textit{based on common, narrowly defined goals} was seen as the path forward to gain adoption and engagement. 
\xquote{What physical things do they get from it? ...By joining this, what does it make? How does it make their job easier or better? If you can't answer those, then they're not going to be there just because you build something.}{P12, Service Provider (Retired law enforcer)}

As depicted in the above quote, across all stakeholder groups, participants wanted to see clear goals and meaningful benefits from their involvement. This consensus points towards the necessity for streamlined engagement strategies that promote valuable contributions through specific and focused goals. Such adjustments could foster a sense of belonging and worth among members, enhancing overall collaboration and effectiveness in community initiatives. 

\textbf{Non-partisan leadership is necessary to define and achieve those common goals.} For a community to organically grow, the stakeholders expected that community leadership would define the goal(s) the community works on. This was seen as providing clear community goals leading to accountability, clear messaging on membership fit, and expected contribution levels. While participants suggested that the community should work, and build membership on a common goal, they noted that they and others are busy and thus guard access to their time. Expecting them to submit new project proposal ideas or heavily contribute to projects was therefore questionable. However, they also expressed a sense of responsibility and altruism in that they were willing to find the time to put towards the cause if their assignments could be explicitly stated, and if they were asked directly to fill a need. From a community-wide perspective, participants expect the community software implementation to cater to this by allowing them to contribute their knowledge asynchronously through knowledge bases and question+answer formats. This would allow members to passively contribute to the community while not placing strict demands on their already constrained time.
\xquote{I, personally would love you know, a world-leading community that summarized the peer-reviewed research and had that in ways that were very accessible... That kind of, you know, shows the benefits and shows the drawbacks, and so on, would be helpful.}{P29, Industry (Entrepreneur)}

The interplay between the desire to help when directly requested to fill a need and apprehension to work on strategies to bring stakeholders together also raised the concept of accountability. While stakeholders wanted to be credited for their work and credit others as shown earlier in the results, they also expressed the need to hold someone accountable in less than favorable conditions. This was expressed in two ways. Firstly, participants viewed open innovation initiatives with skepticism regarding the longevity and availability of ongoing maintenance of the projects. Skepticism also was shown toward the perceived lack of being able to hold the open innovation project accountable for poor performance. Secondly, participants, mostly in the service provider group, expressed the need for a resource to turn to when they need the resource that has the answers to their issues. The participants expressed a need to hold someone accountable to have the knowledge they don't and for that person/organization to keep abreast with the challenges youth face. Access to this knowledge was seen as being the value of community participation.
In summary, participants suggested that for a multi-stakeholder community to be successful, there should be well-defined common goals to pursue together. To define common goals and move toward those goals, participants wanted non-partisan leadership to lead the goal-setting and collaboration toward those common goals. 

\section{Discussion}
We present a discussion based on the major takeaways of our study and implications for building a multi-stakeholder youth online safety community to advance the state-of-the-art in sociotechnical interventions for youth protection. We end with the limitations of our study and possible avenues for future inquiry.

\subsection{Resolving Tensions (RQ1) to Converge on a Path Forward (RQ2)}

In this study, participants were presented with a bare-bones online community prototype focused on making youth online safety technology with an example of such technology shown as a design probe. Overall, participants exhibited strong positive feelings toward the need for youth to be protected from online risk and felt that youth online risk issues would not fix themselves. At the same time, we observe that collaborative efforts in the multistakeholder youth online safety community have to overcome tensions that exist between the stakeholders. In this section, we discuss such tensions and map potential solutions to converge on a path forward.

\subsubsection{Clear and actionable goals are needed for targeted involvement} \edit{In RQ1, we investigated tensions that arise when youth online safety stakeholders conceptualize an open innovation community. Tensions surfaced between each stakeholder group when examining operating as a community, the contributions they can make to the community, and the different outcomes they expect the community to deliver. When viewed under the perspective of Olson's four-dimensional framework~\cite{Olson2000}, our findings align with previous research in that finding common ground, appropriately dividing work across teams, and paying close attention to the collaboration readiness of team members is pivotal to the success of youth online safety projects. Youth online safety stakeholders all meet Olson's dimensions of common ground (partially), and collaboration readiness \cite{Olson2000} in that they earnestly desire to help youth have safe online interactions, and transition safely to adulthood, even though their approaches and agendas differ. The divergence of approaches and agendas leads to only partial attainment of common ground and possible distrust.}

While the stakeholders respect each other in a collaborative setting, the preceding dynamic created an environment of distrust where efforts in youth online safety are often viewed with skepticism especially when the aims of the efforts are not clearly understood by key stakeholders~\cite{fisk2016framing}. Therefore, successful youth online safety communities must utilize a method to reconcile these differing end goals with pinpointed contribution agendas. Clear goal setting, falling under Olson's nature of work dimension~\cite{Olson2000}, with targets that can be actioned on and achieved \edit{\textit{collaboratively}}, is necessary for multistakeholder collaboration. \edit{This approach would represent a movement away from the duplicative implementation silos uncovered in previous youth online safety studies~\cite{caddle2024}, toward a more unified approach to solutions fostered by the community}. To this end, community outcomes could include both partial solutions (per researchers) that help generate empirical evidence for understanding the problem and evaluative metrics towards ensuring the efficacy of the approach, as well as end-to-end solutions (per industry professionals) that are commercially viable for broad dissemination. Yet, in both cases, these partial and full-cycle solutions need to be beneficial to \edit{primary stakeholders,} youth and parents (per social service providers), to the extent that they raise awareness and empower youth and parents to achieve beneficial online outcomes. \edit{Therefore, successful youth online safety communities must utilize a method to reconcile these differing end goals with pinpointed contribution agendas. Good management and leadership in collaborations is a promising path toward this reconciliation \cite{Olson2009} which we will discuss in detail below.}


\subsubsection{Leadership is needed for collaborative outcomes} \edit{In RQ2, we investigated what could be done to resolve the tensions identified by participants. Our findings point towards the need for good management and decision-making to alleviate the tensions in which human mediation has been shown to aid in resolving tensions in the technology working groups \cite{Olson2009}. This is two-fold, in that they can 1) work as brokers to translate business needs to technology requirements, and interpret technology outputs back to business speak \cite{Zhang2020}, and 2) resolve perceived (or actual) communication and scheduling issues in hierarchical structures, even across teams \cite{Nomura2008}. These findings echo findings by Olson \cite{Olson2009} where fair, transparent, leadership was found to be necessary for successful remote collaboration. While our findings mirror the findings of multidisciplinary stakeholder studies executed under different contexts, the high-stakes area of youth online safety demands that we not only look at these findings from a scholarly perspective but also one of practical implementation. Previous multistakeholder internet governance initiatives have received criticism and stakeholder withdrawal for using approaches that do not uphold the principles of the internet (i.e., transparency, inclusion, decentralized \& distributed methodologies of implementation and governance, accountability, bottom-up focus, openness) ~\cite{iab2014,isoc2014,wiki:ICANN2022}. These clashing efforts and concerns are seen as devolving discussions into debating the use of top-down vs bottom-up methods with limited actionable outcomes~\cite{buckridge2024}, pitting stakeholders against each other, reminiscent of the need to find common ground. } The preceding discussion sets the stage for what is necessary for collaboration in youth online safety communities: clear vision, accountability, respect, and transparency. This now brings our attention to the outcomes of collaboration. Stakeholders envisioned the end goals of collaborative efforts through the lens of how participation helps them achieve the goals of their stakeholder role. These outcomes are 1) quality profitable products for industry, 2) publishable insights for researchers, and 3) decreased feelings of isolation for service providers. While previous literature has discussed desired outcomes through the perspectives of social service providers~\cite{caddle2022}, and also suggested working as a community to create youth online safety technologies~\cite{caddle2024}, what remained open was defining a path forward to integrate these seemingly divergent outcomes to foster collaboration. Participants in our study likewise did not envision concretely what end goals the collaborative efforts should have other than those presented through their \textit{own} stakeholder group outcomes. Rather, the expectation arising from the overall analysis was that community leaders would set the agenda, timelines, and outcomes, such that the stakeholders could either opt into an effort of their choosing or be directly invited to participate in particular efforts based on the leaders knowing of their skillset. \edit{In the next section, we discuss what unique leadership is needed in multistakeholder collaboration to promote youth online safety. }


\begin{figure}[!htbp]
\centering

\includegraphics[scale=0.40]{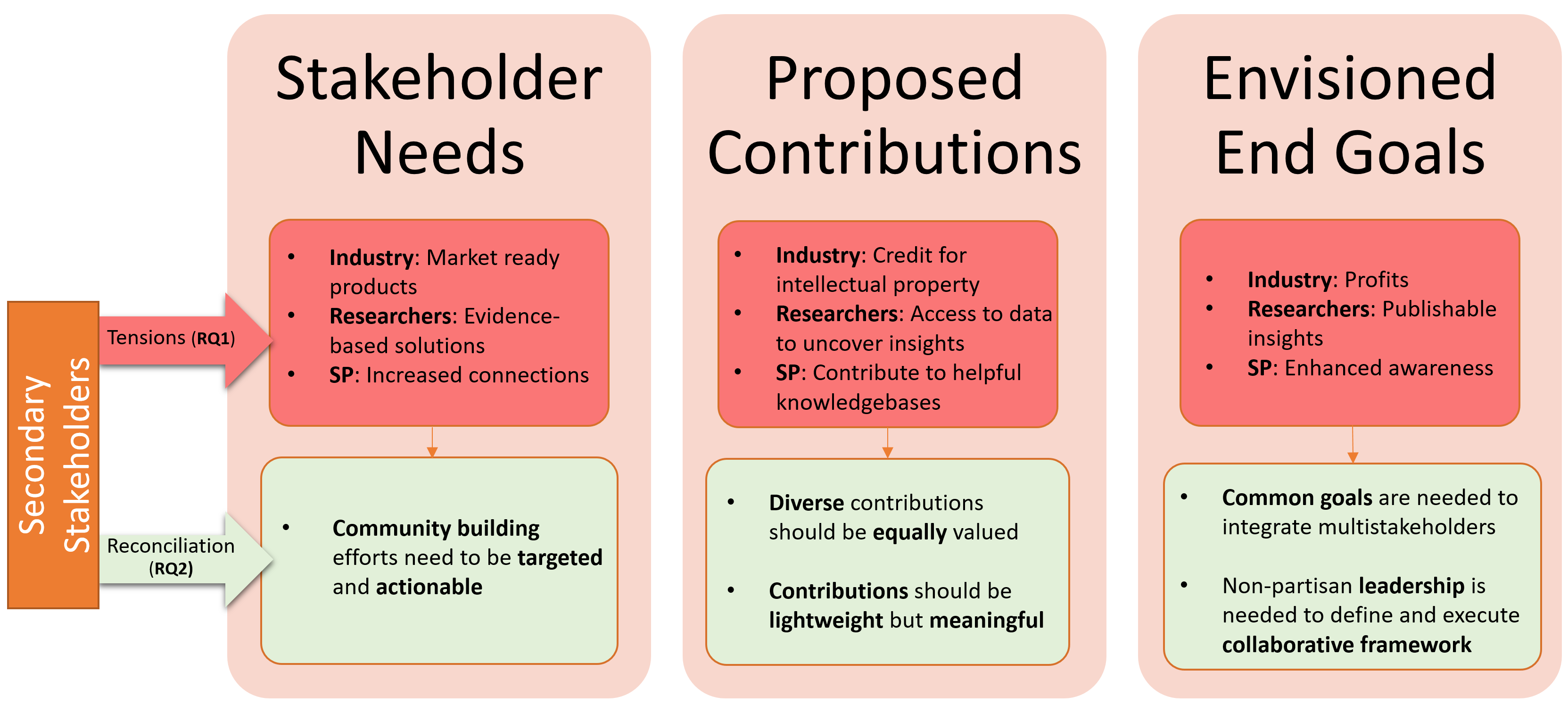}
\caption{Mapping stakeholder tensions to identified solutions}
\Description[Mapping stakeholder tensions to identified solutions.]{Community and contribution tensions can be resolved by providing targeted, actionable goals, and ensuring expected contributions are easy to deliver while being equally valued, respectively. A solution to reconcile collaborative outcomes is not provided.}
\label{fig:study3outcomes}
\end{figure}
\FloatBarrier

\subsubsection{Diverse opinions should be respected towards inclusive goals setting} \edit{The culture of leadership in youth online safety collaboration must be one that encourages cooperation and transparency but also respects the competition that arises when working with industry partners. } In deciding how goals should be formulated for the community of stakeholders, participants envisioned streamlined methods to support each other without jeopardizing their main stakeholder role. This stance puts each stakeholder group at odds with each other as furthering a goal in one stakeholder group might necessitate infringing on the goals of others. For instance, 
the goal of publishing insights puts researchers at odds with service providers who often would prefer to not get youth in trouble through investigations knowing that researchers are mandated reporters when child harm is uncovered~\cite{razi2022instagram, caddle2022}. In efforts to be first to market, industry professionals often want to suppress insights that researchers might want to share to be first to publish. Service providers find themselves on the frontlines facing youth, on the receiving end of industry products and their outcomes good or bad, and more recently, on the research end as participants sought out their unique perspective on youth online safety efforts~\cite{caddle2022,caddle2024}. 
In this environment, a balance is needed that would allow all stakeholders to achieve their goals while working in their various realms. As the literature points out, the intrinsic and extrinsic needs of the stakeholder groups need to be addressed and met for collaboration to be successful~\cite{shibuya2009,ryan2000self, deci1972intrinsic, ryan2020intrinsic}. 

\edit{Mitigating these tensions calls for methodologies that allow for inclusive contributions where the voices and opinions of each individual or stakeholder group are respected and used to chart the paths of community outcomes. This is especially important as the youth online safety space encompasses distinct stakeholder groups with many voices \cite{caddle2024}. In addition to physical distance, at times, global members of the youth online safety community have differing opinions on how the community should move forward, further compounded by the differing cultures in which they reside~\cite{Olson2009}. These physical and conceptual distances between these groups can make developing trust difficult and necessitate intentional actions to build trust \cite{Olson2009}. We next discuss how establishing trust can be achieved.}

\subsection{A Path Towards Open Innovation with Trusted Leadership and Shared and Governance for Facilitating a Youth Online Safety Community}
\edit{While secondary stakeholders in youth online safety see value in contemporary solutions in the field, the overarching tone is the need to go further than solutions that mainly focus on education, research, and policy while respecting the views and needs of all stakeholders. To work together, however, they agree that the collaborative community should \textit{outline a goal they appreciate} (common-ground \cite{Olson2000}). This goal needs to be different from contemporary efforts, which stakeholders see as ``business-as-usual'' and not making the needed impact in the field as per their perspective. The apprehension to take on the initial responsibility of summarizing efforts in the field and setting agendas exposes a need for an effective organization with non-partisan leadership to unify secondary stakeholders, some of who may be competitors in business.}

In this sense, the participants in our study reflected an implicit but strong alignment with open innovation principles. The consensus among stakeholders leaned towards a model governed by a central authority to manage contributions and communication effectively, ensuring quality and relevance, one of the key principles of open innovation~\cite{euchner2013}. 
\edit{The unique open innovation principle we found from the youth online safety community was that } participants highlighted their expectations for leadership that not only embodies expertise and unbiased decision-making but also fosters an inclusive environment where diverse contributions are recognized and valued. 
This preference was noted without significant concern over potential biases this authority might bring, reflecting a general trust in the envisioned leadership's qualifications \edit{toward social good: digital well-being of youth}.
This trust in leadership was seen as essential for ensuring that the principles of transparency, accountability, and equitable participation are upheld, thereby enabling a thriving ecosystem of open innovation that can navigate the complexities and challenges of stakeholder engagement among diverse groups. 

The fields of computer-supported cooperative work (CSCW) and human-computer interaction (HCI) have long considered questions of governance in terms of, \textit{inter alia}, roles, norms, standards, and policies as a means to ensure the protection of users' data, privacy, and psyches \cite{ackerman2000intellectual}. Scholar-practitioners recognize the need for an interdisciplinary approach that brings diverse constituencies together representing policy, practice, and design \cite{jackson2014policy}. For instance, prior work has contemplated designing well-governed innovation processes across a wide variety of areas ranging from digital healthcare \cite{munson2013sociotechnical} and the intersection of healthcare and social media \cite{vlachokyriakos2021research}, to rural and economically developing infrastructure \cite{ruller2021technology, saha2022towards} and refugees in war zones~\cite{fisher2022people}. However, despite the recognition that youth are a vulnerable user group, especially those in at-risk situations \cite{badillo2017abandoned}, the interplay of youth online safety with governance and open innovation has been rarely addressed in the CSCW and broader SIGCHI communities. This study is an important initial step in the necessary investigation into the cross-disciplinary needs of the youth online safety community regarding collaborative opportunities to achieve a more cohesive approach. By offering a unifying solution that brings together the voices of multiple stakeholders to foster dialogue and innovation related to youth online safety, we can provide evidence-based guidelines to promote collaboration using governance and open innovation approaches.

\subsubsection{Guidelines for Open Innovation with Shared Governance for Youth Online Safety} 
Considering the dynamics, the youth online safety community can consider applying an open innovation model. This model would involve a central coordinating body responsible for overarching standards and ensuring consistency in communication and contribution quality.  
Therefore, we suggest the following action items to instantiate the youth online safety community:

\begin{itemize}
    \item \textbf{Establish a governance committee that provides non-partisan leadership}: To combat the fragmentation of the youth online safety community, clear leadership with collectively developed governance processes will be developed. This will engage and encourage stakeholders to participate in open innovation solutions, even when those solutions were initially proprietary \cite{omahony2022proprietary}.
    \item \textbf{Recruit stakeholder representatives to provide a bridge between central leadership and individual working groups}: To implement a collaborative framework, each stakeholder group will form a separate node (i.e., federation) with autonomy over specific domains of expertise. 
    Each node would operate semi-independently but coordinate closely through a central council, such as a steering committee, composed of representatives from each group~\cite{harris2021infrastructuring}. This council would oversee the integration of efforts and ensure that the consortium’s strategies are aligned with the overarching goal of improving online safety for youth.
    \item \textbf{Implement working groups to address specific tasks}: To encourage and preserve the spirit of open innovation, the central body will work in concert with smaller, semi-autonomous groups or committees focused on specific project areas. This approach would incorporate feedback from participants who want to work on discrete topics (i.e., bias in algorithms) or tasks (i.e. technology creation) while balancing the need for centralized oversight with the flexibility and inclusivity essential for a thriving open innovation ecosystem. This model reflects the contemporary people-centric ethic seen in CSCW research and practice~\cite{ackerman2013sharing}.
    \item \textbf{Build integrated and independent online platforms}: To facilitate collaboration among stakeholders, integrated and independent community platforms will be created. This will help alleviate time constraints, data usage concerns, and tension toward platform leadership, all of which are common challenges in the CSCW context~\cite{jiang2017collaborative}.   
\end{itemize}

\edit{
\subsubsection{Guidelines for Implementation of Youth Online Safety Steering Committee}
\hyperref[fig:committee]{Figure \ref{fig:committee}} illustrates a five-step conceptual framework for the creation of a multistakeholder network to enhance youth online safety through AI and software integration. First, a multidisciplinary steering committee, comprising representatives from academia, industry, social services, government, education, parent and youth groups, and AI experts, would define the mission, objectives, and governance structure. The committee would then develop an open-source platform for collaborative contributions, enabling stakeholders to share AI models, methodologies, and resources. Specialized working groups would address areas such as AI development, policy advocacy, education, and ethics, with clear tasks and milestones. An ethical task force would implement guidelines prioritizing data privacy, consent, and responsible technology use, ensuring transparency through public access and peer validation. Finally, the consortium would pilot AI tools in real-world settings, refine them based on feedback, and expand globally, aligning policy initiatives with technological advancements through collaboration with regulatory entities.

\begin{figure}[t]
\centering

\includegraphics[scale=2]{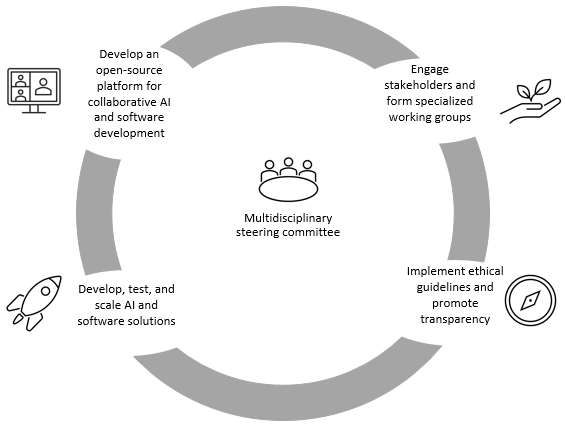}
\caption{Multistakeholder steering committee framework}
\Description{This figure shows proposed implementation guidelines for an organization using open innovation to promote collaboration across stakeholder groups. A non-partisan governance committee provides the governance needed to ensure collaboration across the stakeholder representative groups who in turn manage working groups created to produce particular outcomes. The stages shown are iterative. The steering committee can start at any stage depending on the needs of the particular task at hand.}
\label{fig:committee}
\end{figure}

Although this framework has the potential to enhance youth online safety through a multistakeholder approach, feasibility challenges remain. Coordinating diverse representatives from academia, industry, social services, government, education, and parent and youth groups requires careful mediation and engagement to align differing priorities and operational cultures. Consulting psychologists with expertise in group dynamics and organizational leadership could support governance by fostering collaboration, balancing decision-making power, and addressing issues like conflicting interests and communication barriers \cite{Lowman2016}  \cite{Deutsch2001}. 
Sustained funding will rely on leveraging member networks to build partnerships with governments, nonprofits, and the private sector. Ethical and legal complexities, particularly around data privacy, consent, and youth protection, require a flexible, globally applicable ethical framework. A legal and ethics advisory task force can help navigate these challenges. Technological hurdles, including open-source security and peer validation, may necessitate dedicated teams, while scalability will depend on tailoring solutions to regional variations. By addressing these challenges with a structured yet adaptable approach, the committee can create a meaningful and sustainable initiative.
}

\subsection{Limitations and Future Work}
We acknowledge that our study has several limitations that should be addressed in future work. All of our study participants were from the U.S. or operated businesses that cater to North American populations. Caution should therefore be taken in generalizing the perspectives presented as the sample is not representative of the worldwide population. Focusing primarily on U.S.-based stakeholders for this study was a conscious decision aimed at bringing parity to the youth online safety collaborations already well-established in other countries. However, it is important that future work take a multi-national multi-stakeholder perspective to tie all of these efforts together for a synergistic global strategy for youth online safety in digital spaces. 

This study also did not capture the views of the primary stakeholders, parents, and youth, and how to integrate them directly as contributors to the community. The views of primary stakeholders in youth online safety have been well-established in the literature. Only recently, the views of secondary stakeholders have come into focus~\cite{caddle2022,caddle2024} and our work is one such effort to address an under-explored area. Still, there is work to be done to bring the primary and secondary stakeholders together under the context of collaborative youth online safety design. Future work should focus on integrating the perspectives of these distinct stakeholder groups.

\section{Conclusion}
Secondary stakeholders in youth online safety view efforts in the area as important regardless of the size of the organization undertaking the effort. However, fragmentation exists regarding their perspectives on how protections should be designed and who should shoulder the major responsibility for implementing those protections. While overall the prevailing thought is that everyone must play a role in this effort, the fragmentation creates an environment where critiquing the efforts of others is easy, but working with them to enhance solutions is hard. This frustrates secondary stakeholders, leaving them feeling fatigued and isolated while also negatively impacting the primary stakeholders in youth online safety, the youth. Secondary stakeholders exhibit a willingness to work together in protection efforts if a structure exists that allows them to contribute their knowledge without encroaching on proprietary material. To capitalize on this, and combat the fragmentation, we propose the creation of a non-partisan organization operating under an open innovation methodology to unify youth online safety efforts, thereby working with youth-facing internet companies and the public to provide solutions. This organization can aid in the assimilation of public voices into design patterns and the creation of open-standards for youth-facing internet companies for more cohesive protections.

\begin{acks}
This research is partially supported by the U.S. National Science Foundation under grants \censor{IIP-2329976} and \censor{IIS-2333207} and by the William T. Grant Foundation grant \censor{\#187941}. Any opinions, findings, conclusions, or recommendations expressed in this paper do not necessarily reflect the views of the sponsors. We would also like to thank \censor{Ashwaq Alsoubai, Andrew Niser, Abdulmalik Alluhidan, and Zainab Agha, from Vanderbilt University} for their assistance.
\end{acks}

\bibliographystyle{ACM-Reference-Format}
\bibliography{00-references}


\begin{thebibliography}{89}


\ifx \showCODEN    \undefined \def \showCODEN     #1{\unskip}     \fi
\ifx \showDOI      \undefined \def \showDOI       #1{#1}\fi
\ifx \showISBNx    \undefined \def \showISBNx     #1{\unskip}     \fi
\ifx \showISBNxiii \undefined \def \showISBNxiii  #1{\unskip}     \fi
\ifx \showISSN     \undefined \def \showISSN      #1{\unskip}     \fi
\ifx \showLCCN     \undefined \def \showLCCN      #1{\unskip}     \fi
\ifx \shownote     \undefined \def \shownote      #1{#1}          \fi
\ifx \showarticletitle \undefined \def \showarticletitle #1{#1}   \fi
\ifx \showURL      \undefined \def \showURL       {\relax}        \fi
\providecommand\bibfield[2]{#2}
\providecommand\bibinfo[2]{#2}
\providecommand\natexlab[1]{#1}
\providecommand\showeprint[2][]{arXiv:#2}

\bibitem[Ackerman(2000)]%
        {ackerman2000intellectual}
\bibfield{author}{\bibinfo{person}{Mark~S Ackerman}.} \bibinfo{year}{2000}\natexlab{}.
\newblock \showarticletitle{The intellectual challenge of CSCW: The gap between social requirements and technical feasibility}.
\newblock \bibinfo{journal}{\emph{Human--Computer Interaction}} \bibinfo{volume}{15}, \bibinfo{number}{2-3} (\bibinfo{year}{2000}), \bibinfo{pages}{179--203}.
\newblock


\bibitem[Ackerman et~al\mbox{.}(2013)]%
        {ackerman2013sharing}
\bibfield{author}{\bibinfo{person}{Mark~S Ackerman}, \bibinfo{person}{Juri Dachtera}, \bibinfo{person}{Volkmar Pipek}, {and} \bibinfo{person}{Volker Wulf}.} \bibinfo{year}{2013}\natexlab{}.
\newblock \showarticletitle{Sharing knowledge and expertise: The CSCW view of knowledge management}.
\newblock \bibinfo{journal}{\emph{Computer Supported Cooperative Work (CSCW)}}  \bibinfo{volume}{22} (\bibinfo{year}{2013}), \bibinfo{pages}{531--573}.
\newblock


\bibitem[Agha et~al\mbox{.}(2023)]%
        {agha2023strike}
\bibfield{author}{\bibinfo{person}{Zainab Agha}, \bibinfo{person}{Karla Badillo-Urquiola}, {and} \bibinfo{person}{Pamela~J Wisniewski}.} \bibinfo{year}{2023}\natexlab{}.
\newblock \showarticletitle{" Strike at the Root": Co-designing Real-Time Social Media Interventions for Adolescent Online Risk Prevention}.
\newblock \bibinfo{journal}{\emph{Proceedings of the ACM on Human-Computer Interaction}} \bibinfo{volume}{7}, \bibinfo{number}{CSCW1} (\bibinfo{year}{2023}), \bibinfo{pages}{1--32}.
\newblock


\bibitem[Ahmed and Guha(2012)]%
        {Ahmed2012}
\bibfield{author}{\bibinfo{person}{Syed~Ishtiaque Ahmed} {and} \bibinfo{person}{Shion Guha}.} \bibinfo{year}{2012}\natexlab{}.
\newblock \showarticletitle{Distance matters: an exploratory analysis of the linguistic features of Flickr photo tag metadata in relation to impression management}. In \bibinfo{booktitle}{\emph{Proceedings of the 2nd ACM SIGMOD Workshop on Databases and Social Networks}} (Scottsdale, Arizona) \emph{(\bibinfo{series}{DBSocial '12})}. \bibinfo{publisher}{Association for Computing Machinery}, \bibinfo{address}{New York, NY, USA}, \bibinfo{pages}{7–12}.
\newblock
\showISBNx{9781450314954}
\urldef\tempurl%
\url{https://doi.org/10.1145/2304536.2304538}
\showDOI{\tempurl}


\bibitem[Ahn et~al\mbox{.}(2019)]%
        {ahn2019leveraging}
\bibfield{author}{\bibinfo{person}{Joon~Mo Ahn}, \bibinfo{person}{Nadine Roijakkers}, \bibinfo{person}{Riccardo Fini}, {and} \bibinfo{person}{Letizia Mortara}.} \bibinfo{year}{2019}\natexlab{}.
\newblock \bibinfo{title}{Leveraging open innovation to improve society: past achievements and future trajectories}.
\newblock , \bibinfo{numpages}{267--278}~pages.
\newblock


\bibitem[Akter et~al\mbox{.}(2022)]%
        {akter2022parental}
\bibfield{author}{\bibinfo{person}{Mamtaj Akter}, \bibinfo{person}{Amy~J Godfrey}, \bibinfo{person}{Jess Kropczynski}, \bibinfo{person}{Heather~R Lipford}, {and} \bibinfo{person}{Pamela~J Wisniewski}.} \bibinfo{year}{2022}\natexlab{}.
\newblock \showarticletitle{From Parental Control to Joint Family Oversight: Can Parents and Teens Manage Mobile Online Safety and Privacy as Equals?}
\newblock \bibinfo{journal}{\emph{Proceedings of the ACM on Human-Computer Interaction}} \bibinfo{volume}{6}, \bibinfo{number}{CSCW1} (\bibinfo{year}{2022}), \bibinfo{pages}{1--28}.
\newblock


\bibitem[Alluhidan et~al\mbox{.}(2024)]%
        {alluhidan2024teen}
\bibfield{author}{\bibinfo{person}{Abdulmalik Alluhidan}, \bibinfo{person}{Mamtaj Akter}, \bibinfo{person}{Ashwaq Alsoubai}, \bibinfo{person}{Jinkyung~Katie Park}, {and} \bibinfo{person}{Pamela Wisniewski}.} \bibinfo{year}{2024}\natexlab{}.
\newblock \showarticletitle{Teen Talk: The Good, the Bad, and the Neutral of Adolescent Social Media Use}.
\newblock \bibinfo{journal}{\emph{Proceedings of the ACM on Human-Computer Interaction}} \bibinfo{volume}{8}, \bibinfo{number}{CSCW2} (\bibinfo{year}{2024}), \bibinfo{pages}{1--35}.
\newblock


\bibitem[Alsoubai et~al\mbox{.}(2022a)]%
        {alsoubai2022}
\bibfield{author}{\bibinfo{person}{Ashwaq Alsoubai}, \bibinfo{person}{Xavier~V. Caddle}, \bibinfo{person}{Ryan Doherty}, \bibinfo{person}{Alexandra~Taylor Koehler}, \bibinfo{person}{Estefania Sanchez}, \bibinfo{person}{Munmun De~Choudhury}, {and} \bibinfo{person}{Pamela~J. Wisniewski}.} \bibinfo{year}{2022}\natexlab{a}.
\newblock \showarticletitle{MOSafely, Is That Sus? A Youth-Centric Online Risk Assessment Dashboard}. In \bibinfo{booktitle}{\emph{Companion Publication of the 2022 Conference on Computer Supported Cooperative Work and Social Computing}} (Virtual Event, Taiwan) \emph{(\bibinfo{series}{CSCW'22 Companion})}. \bibinfo{publisher}{Association for Computing Machinery}, \bibinfo{address}{New York, NY, USA}, \bibinfo{pages}{197–200}.
\newblock
\showISBNx{9781450391900}
\urldef\tempurl%
\url{https://doi.org/10.1145/3500868.3559710}
\showDOI{\tempurl}


\bibitem[Alsoubai et~al\mbox{.}(2022b)]%
        {alsoubai2022friends}
\bibfield{author}{\bibinfo{person}{Ashwaq Alsoubai}, \bibinfo{person}{Jihye Song}, \bibinfo{person}{Afsaneh Razi}, \bibinfo{person}{Nurun Naher}, \bibinfo{person}{Munmun De~Choudhury}, {and} \bibinfo{person}{Pamela~J Wisniewski}.} \bibinfo{year}{2022}\natexlab{b}.
\newblock \showarticletitle{From'Friends with Benefits' to'Sextortion:'A Nuanced Investigation of Adolescents' Online Sexual Risk Experiences}.
\newblock \bibinfo{journal}{\emph{Proceedings of the ACM on Human-Computer Interaction}} \bibinfo{volume}{6}, \bibinfo{number}{CSCW2} (\bibinfo{year}{2022}), \bibinfo{pages}{1--32}.
\newblock


\bibitem[Anderson et~al\mbox{.}(2023)]%
        {Anderson2023}
\bibfield{author}{\bibinfo{person}{Monica Anderson}, \bibinfo{person}{Michelle Faverio}, {and} \bibinfo{person}{Gottfried Jeffrey}.} \bibinfo{year}{2023}\natexlab{}.
\newblock \bibinfo{title}{Teens, {Social} {Media} \& {Technology} 2023 {\textbar} {Pew} {Research} {Center}}.
\newblock
\newblock
\urldef\tempurl%
\url{https://www.pewresearch.org/internet/2023/12/11/teens-social-media-and-technology-2023/}
\showURL{%
\tempurl}


\bibitem[Badillo-Urquiola et~al\mbox{.}(2024)]%
        {badillo2024towards}
\bibfield{author}{\bibinfo{person}{Karla Badillo-Urquiola}, \bibinfo{person}{Zainab Agha}, \bibinfo{person}{Denielle Abaquita}, \bibinfo{person}{Scott~B Harpin}, {and} \bibinfo{person}{Pamela~J Wisniewski}.} \bibinfo{year}{2024}\natexlab{}.
\newblock \showarticletitle{Towards a Social Ecological Approach to Supporting Caseworkers in Promoting the Online Safety of Youth in Foster Care}.
\newblock \bibinfo{journal}{\emph{Proceedings of the ACM on Human-Computer Interaction}} \bibinfo{volume}{8}, \bibinfo{number}{CSCW1} (\bibinfo{year}{2024}), \bibinfo{pages}{1--28}.
\newblock


\bibitem[Badillo-Urquiola et~al\mbox{.}(2017)]%
        {badillo2017abandoned}
\bibfield{author}{\bibinfo{person}{Karla Badillo-Urquiola}, \bibinfo{person}{Scott Harpin}, {and} \bibinfo{person}{Pamela Wisniewski}.} \bibinfo{year}{2017}\natexlab{}.
\newblock \showarticletitle{Abandoned but not forgotten: Providing access while protecting foster youth from online risks}. In \bibinfo{booktitle}{\emph{Proceedings of the 2017 Conference on Interaction Design and Children}}. \bibinfo{pages}{17--26}.
\newblock


\bibitem[Badillo-Urquiola et~al\mbox{.}(2020)]%
        {Badillo_children_2020}
\bibfield{author}{\bibinfo{person}{Karla Badillo-Urquiola}, \bibinfo{person}{Afsaneh Razi}, \bibinfo{person}{Jan Edwards}, {and} \bibinfo{person}{Pamela Wisniewski}.} \bibinfo{year}{2020}\natexlab{}.
\newblock \showarticletitle{Children's Perspectives on Human Sex Trafficking Prevention Education}. In \bibinfo{booktitle}{\emph{Companion of the 2020 ACM International Conference on Supporting Group Work}}. \bibinfo{pages}{123--126}.
\newblock


\bibitem[Baldry et~al\mbox{.}(2017)]%
        {Baldry_school_2017}
\bibfield{author}{\bibinfo{person}{Anna~Costanza Baldry}, \bibinfo{person}{David~P Farrington}, {and} \bibinfo{person}{Anna Sorrentino}.} \bibinfo{year}{2017}\natexlab{}.
\newblock \showarticletitle{School bullying and cyberbullying among boys and girls: Roles and overlap}.
\newblock \bibinfo{journal}{\emph{Journal of Aggression, Maltreatment \& Trauma}} \bibinfo{volume}{26}, \bibinfo{number}{9} (\bibinfo{year}{2017}), \bibinfo{pages}{937--951}.
\newblock


\bibitem[Braun et~al\mbox{.}(2019)]%
        {Braun_2019}
\bibfield{author}{\bibinfo{person}{Virginia Braun}, \bibinfo{person}{Victoria Clarke}, \bibinfo{person}{Nikki Hayfield}, {and} \bibinfo{person}{Gareth Terry}.} \bibinfo{year}{2019}\natexlab{}.
\newblock \bibinfo{booktitle}{\emph{Thematic Analysis}}.
\newblock \bibinfo{publisher}{Springer Singapore}, \bibinfo{address}{Singapore}, \bibinfo{pages}{843--860}.
\newblock
\showISBNx{978-981-10-5251-4}
\urldef\tempurl%
\url{https://doi.org/10.1007/978-981-10-5251-4_103}
\showDOI{\tempurl}


\bibitem[Buckridge(2024)]%
        {buckridge2024}
\bibfield{author}{\bibinfo{person}{Chris Buckridge}.} \bibinfo{year}{2024}\natexlab{}.
\newblock \bibinfo{booktitle}{\emph{Why We Need Multistakeholder Internet Governance}}.
\newblock Council of European National Top-level Domain Registries.
\newblock
\urldef\tempurl%
\url{https://www.centr.org/component/flexicontent/download/10932/7952/41.html}
\showURL{%
Retrieved May 30, 2024 from \tempurl}


\bibitem[Caddle et~al\mbox{.}(2024)]%
        {caddle2024}
\bibfield{author}{\bibinfo{person}{Xavier Caddle}, \bibinfo{person}{Jinkyung~Katie Park}, {and} \bibinfo{person}{Pamela~J. Wisniewski}.} \bibinfo{year}{2024}\natexlab{}.
\newblock \showarticletitle{A Stakeholders' Analysis of the Sociotechnical Approaches for Protecting Youth Online}. In \bibinfo{booktitle}{\emph{Advances in Information and Communication}}, \bibfield{editor}{\bibinfo{person}{Kohei Arai}} (Ed.). \bibinfo{publisher}{Springer Nature Switzerland}, \bibinfo{address}{Cham}, \bibinfo{pages}{587--616}.
\newblock
\showISBNx{978-3-031-54053-0}
\urldef\tempurl%
\url{https://doi.org/10.1007/978-3-031-54053-0_40}
\showDOI{\tempurl}


\bibitem[Caddle et~al\mbox{.}(2022)]%
        {caddle2022}
\bibfield{author}{\bibinfo{person}{Xavier~V. Caddle}, \bibinfo{person}{Nurun Naher}, \bibinfo{person}{Zachary~P. Miller}, \bibinfo{person}{Karla Badillo-Urquiola}, {and} \bibinfo{person}{Pamela~J. Wisniewski}.} \bibinfo{year}{2022}\natexlab{}.
\newblock \showarticletitle{Duty to Respond: The Challenges Social Service Providers Face When Charged with Keeping Youth Safe Online}.
\newblock \bibinfo{journal}{\emph{Proc. ACM Hum.-Comput. Interact.}} \bibinfo{volume}{7}, \bibinfo{number}{GROUP}, Article \bibinfo{articleno}{6} (\bibinfo{date}{dec} \bibinfo{year}{2022}), \bibinfo{numpages}{35}~pages.
\newblock
\urldef\tempurl%
\url{https://doi.org/10.1145/3567556}
\showDOI{\tempurl}


\bibitem[Chesbrough et~al\mbox{.}(2014)]%
        {chesbrough2014open}
\bibfield{author}{\bibinfo{person}{Henry Chesbrough}, \bibinfo{person}{Alberto Di~Minin}, {et~al\mbox{.}}} \bibinfo{year}{2014}\natexlab{}.
\newblock \showarticletitle{Open social innovation}.
\newblock \bibinfo{journal}{\emph{New frontiers in open innovation}}  \bibinfo{volume}{16} (\bibinfo{year}{2014}), \bibinfo{pages}{301--315}.
\newblock


\bibitem[Chesbrough et~al\mbox{.}(2006)]%
        {chesbrough2006open}
\bibfield{author}{\bibinfo{person}{Henry Chesbrough}, \bibinfo{person}{Wim Vanhaverbeke}, {and} \bibinfo{person}{Joel West}.} \bibinfo{year}{2006}\natexlab{}.
\newblock \bibinfo{booktitle}{\emph{Open innovation: Researching a new paradigm}}.
\newblock \bibinfo{publisher}{Oxford university press, USA}.
\newblock


\bibitem[Chesbrough(2003)]%
        {chesbrough2003}
\bibfield{author}{\bibinfo{person}{Henry~William Chesbrough}.} \bibinfo{year}{2003}\natexlab{}.
\newblock \bibinfo{booktitle}{\emph{Open innovation: The new imperative for creating and profiting from technology}}.
\newblock \bibinfo{publisher}{Harvard Business Press}.
\newblock


\bibitem[Commission(2022)]%
        {FTA2022combatting}
\bibfield{author}{\bibinfo{person}{Federal~Trade Commission}.} \bibinfo{year}{2022}\natexlab{}.
\newblock \bibinfo{title}{Combatting Online Harms Through Innovation}.
\newblock
\newblock
\urldef\tempurl%
\url{https://www.ftc.gov/system/files/ftc_gov/pdf/Combatting\%20Online\%20Harms\%20Through\%20Innovation\%3B\%20Federal\%20Trade\%20Commission\%20Report\%20to\%20Congress.pdf}
\showURL{%
\tempurl}


\bibitem[Das et~al\mbox{.}(2024)]%
        {das2024comes}
\bibfield{author}{\bibinfo{person}{Maitraye Das}, \bibinfo{person}{Abigale Stangl}, {and} \bibinfo{person}{Leah Findlater}.} \bibinfo{year}{2024}\natexlab{}.
\newblock \showarticletitle{" That comes with a huge career cost:" Understanding Collaborative Ideation Experiences of Disabled Professionals}.
\newblock \bibinfo{journal}{\emph{Proceedings of the ACM on Human-Computer Interaction}} \bibinfo{volume}{8}, \bibinfo{number}{CSCW1} (\bibinfo{year}{2024}), \bibinfo{pages}{1--28}.
\newblock


\bibitem[Davidson et~al\mbox{.}(2014)]%
        {davidson2014}
\bibfield{author}{\bibinfo{person}{Jennifer~L. Davidson}, \bibinfo{person}{Rithika Naik}, \bibinfo{person}{Umme~Ayda Mannan}, \bibinfo{person}{Amir Azarbakht}, {and} \bibinfo{person}{Carlos Jensen}.} \bibinfo{year}{2014}\natexlab{}.
\newblock \showarticletitle{On older adults in free/open source software: reflections of contributors and community leaders}. In \bibinfo{booktitle}{\emph{2014 IEEE Symposium on Visual Languages and Human-Centric Computing (VL/HCC)}}. \bibinfo{pages}{93--100}.
\newblock
\urldef\tempurl%
\url{https://doi.org/10.1109/VLHCC.2014.6883029}
\showDOI{\tempurl}


\bibitem[Davis et~al\mbox{.}(2023)]%
        {davis2023supporting}
\bibfield{author}{\bibinfo{person}{Katie Davis}, \bibinfo{person}{Petr Slovak}, \bibinfo{person}{Rotem Landesman}, \bibinfo{person}{Caroline Pitt}, \bibinfo{person}{Abdullatif Ghajar}, \bibinfo{person}{Jessica~Lee Schleider}, \bibinfo{person}{Saba Kawas}, \bibinfo{person}{Andrea~Guadalupe Perez~Portillo}, {and} \bibinfo{person}{Nicole~S Kuhn}.} \bibinfo{year}{2023}\natexlab{}.
\newblock \showarticletitle{Supporting Teens’ Intentional Social Media Use Through Interaction Design: An exploratory proof-of-concept study}. In \bibinfo{booktitle}{\emph{Proceedings of the 22nd Annual ACM Interaction Design and Children Conference}}. \bibinfo{pages}{322--334}.
\newblock


\bibitem[Deci(1972)]%
        {deci1972intrinsic}
\bibfield{author}{\bibinfo{person}{Edward~L Deci}.} \bibinfo{year}{1972}\natexlab{}.
\newblock \showarticletitle{Intrinsic motivation, extrinsic reinforcement, and inequity.}
\newblock \bibinfo{journal}{\emph{Journal of personality and social psychology}} \bibinfo{volume}{22}, \bibinfo{number}{1} (\bibinfo{year}{1972}), \bibinfo{pages}{113}.
\newblock


\bibitem[Deutsch(2001)]%
        {Deutsch2001}
\bibfield{author}{\bibinfo{person}{Morton Deutsch}.} \bibinfo{year}{2001}\natexlab{}.
\newblock \showarticletitle{Cooperation and conflict resolution: Implications for consulting psychology.}
\newblock \bibinfo{journal}{\emph{Consulting Psychology Journal: Practice and Research}} \bibinfo{volume}{53}, \bibinfo{number}{2} (\bibinfo{year}{2001}), \bibinfo{pages}{76}.
\newblock


\bibitem[Eppinger(2021)]%
        {eppinger2021open}
\bibfield{author}{\bibinfo{person}{Elisabeth Eppinger}.} \bibinfo{year}{2021}\natexlab{}.
\newblock \showarticletitle{How open innovation practices deliver societal benefits}.
\newblock \bibinfo{journal}{\emph{Sustainability}} \bibinfo{volume}{13}, \bibinfo{number}{3} (\bibinfo{year}{2021}), \bibinfo{pages}{1431}.
\newblock


\bibitem[Euchner(2013)]%
        {euchner2013}
\bibfield{author}{\bibinfo{person}{Jim Euchner}.} \bibinfo{year}{2013}\natexlab{}.
\newblock \showarticletitle{The Uses and Risks of Open Innovation}.
\newblock \bibinfo{journal}{\emph{Research-Technology Management}} \bibinfo{volume}{56}, \bibinfo{number}{3} (\bibinfo{year}{2013}), \bibinfo{pages}{49--54}.
\newblock
\urldef\tempurl%
\url{https://doi.org/10.5437/08956308X5603936}
\showDOI{\tempurl}
\showeprint{https://doi.org/10.5437/08956308X5603936}


\bibitem[Feng et~al\mbox{.}(2023)]%
        {feng2023understanding}
\bibfield{author}{\bibinfo{person}{KJ~Kevin Feng}, \bibinfo{person}{Tony~W Li}, {and} \bibinfo{person}{Amy~X Zhang}.} \bibinfo{year}{2023}\natexlab{}.
\newblock \showarticletitle{Understanding collaborative practices and tools of professional UX practitioners in software organizations}. In \bibinfo{booktitle}{\emph{Proceedings of the 2023 CHI Conference on Human Factors in Computing Systems}}. \bibinfo{pages}{1--20}.
\newblock


\bibitem[Fisher(2022)]%
        {fisher2022people}
\bibfield{author}{\bibinfo{person}{Karen~E Fisher}.} \bibinfo{year}{2022}\natexlab{}.
\newblock \showarticletitle{People first, data second: A humanitarian research framework for fieldwork with refugees by war zones}.
\newblock \bibinfo{journal}{\emph{Computer Supported Cooperative Work (CSCW)}} \bibinfo{volume}{31}, \bibinfo{number}{2} (\bibinfo{year}{2022}), \bibinfo{pages}{237--297}.
\newblock


\bibitem[Fisk(2016)]%
        {fisk2016framing}
\bibfield{author}{\bibinfo{person}{Nathan~W Fisk}.} \bibinfo{year}{2016}\natexlab{}.
\newblock \bibinfo{booktitle}{\emph{Framing internet safety: The governance of youth online}}.
\newblock \bibinfo{publisher}{MIT Press}.
\newblock


\bibitem[Freeman and Auster(2012)]%
        {freeman2012values}
\bibfield{author}{\bibinfo{person}{R~Edward Freeman} {and} \bibinfo{person}{Ellen~R Auster}.} \bibinfo{year}{2012}\natexlab{}.
\newblock \showarticletitle{Values, authenticity, and responsible leadership}.
\newblock \bibinfo{journal}{\emph{Responsible leadership}} (\bibinfo{year}{2012}), \bibinfo{pages}{15--23}.
\newblock


\bibitem[Ghapanchi et~al\mbox{.}(2011)]%
        {ghapanchi2011}
\bibfield{author}{\bibinfo{person}{Amir~Hossein Ghapanchi}, \bibinfo{person}{Aybuke Aurum}, {and} \bibinfo{person}{Graham Low}.} \bibinfo{year}{2011}\natexlab{}.
\newblock \showarticletitle{A taxonomy for measuring the success of open source software projects}.
\newblock \bibinfo{journal}{\emph{First Monday}} \bibinfo{volume}{16}, \bibinfo{number}{8} (\bibinfo{date}{Jul.} \bibinfo{year}{2011}).
\newblock
\urldef\tempurl%
\url{https://doi.org/10.5210/fm.v16i8.3558}
\showDOI{\tempurl}


\bibitem[Harris et~al\mbox{.}(2021)]%
        {harris2021infrastructuring}
\bibfield{author}{\bibinfo{person}{Marcelline~R Harris}, \bibinfo{person}{Lisa~A Ferguson}, {and} \bibinfo{person}{Airong Luo}.} \bibinfo{year}{2021}\natexlab{}.
\newblock \showarticletitle{Infrastructuring an organizational node for a federated research and data network: A case study from a sociotechnical perspective}.
\newblock \bibinfo{journal}{\emph{Journal of Clinical and Translational Science}} \bibinfo{volume}{5}, \bibinfo{number}{1} (\bibinfo{year}{2021}), \bibinfo{pages}{e186}.
\newblock


\bibitem[Hou and Wang(2017)]%
        {Hou2017}
\bibfield{author}{\bibinfo{person}{Youyang Hou} {and} \bibinfo{person}{Dakuo Wang}.} \bibinfo{year}{2017}\natexlab{}.
\newblock \showarticletitle{Hacking with NPOs: Collaborative Analytics and Broker Roles in Civic Data Hackathons}.
\newblock \bibinfo{journal}{\emph{Proc. ACM Hum.-Comput. Interact.}} \bibinfo{volume}{1}, \bibinfo{number}{CSCW}, Article \bibinfo{articleno}{53} (\bibinfo{date}{Dec.} \bibinfo{year}{2017}), \bibinfo{numpages}{16}~pages.
\newblock
\urldef\tempurl%
\url{https://doi.org/10.1145/3134688}
\showDOI{\tempurl}


\bibitem[ICANNWiki(2022)]%
        {wiki:ICANN2022}
\bibfield{author}{\bibinfo{person}{ICANNWiki}.} \bibinfo{year}{2022}\natexlab{}.
\newblock \bibinfo{title}{NETmundial --- ICANNWiki{,}}.
\newblock
\newblock
\urldef\tempurl%
\url{https://icannwiki.org/NETmundial}
\showURL{%
\tempurl}
\newblock
\shownote{[Online; accessed 30-May-2024]}.


\bibitem[IGF(2024)]%
        {igf2024}
\bibfield{author}{\bibinfo{person}{IGF}.} \bibinfo{year}{2024}\natexlab{}.
\newblock \bibinfo{booktitle}{\emph{Internet Governance}}.
\newblock Internet Governance Forum.
\newblock
\urldef\tempurl%
\url{https://www.intgovforum.org/en/about}
\showURL{%
Retrieved May 30, 2024 from \tempurl}


\bibitem[{I}nternet~{A}rchitecture {B}oard(2014)]%
        {iab2014}
\bibfield{author}{\bibinfo{person}{{I}nternet~{A}rchitecture {B}oard}.} \bibinfo{year}{2014}\natexlab{}.
\newblock \bibinfo{booktitle}{\emph{IAB statement on the NETmundial Initiative}}.
\newblock IAB.
\newblock
\urldef\tempurl%
\url{https://datatracker.ietf.org/doc/statement-iab-statement-on-the-netmundial-initiative/}
\showURL{%
Retrieved May 30, 2024 from \tempurl}


\bibitem[{I}nternet {S}ociety(2014)]%
        {isoc2014}
\bibfield{author}{\bibinfo{person}{{I}nternet {S}ociety}.} \bibinfo{year}{2014}\natexlab{}.
\newblock \bibinfo{booktitle}{\emph{Internet Society Statement on the NETmundial Initiative}}.
\newblock Internet Society.
\newblock
\urldef\tempurl%
\url{https://www.internetsociety.org/news/press-releases/2014/internet-society-statement-on-the-netmundial-initiative/}
\showURL{%
Retrieved May 30, 2024 from \tempurl}


\bibitem[Jackson et~al\mbox{.}(2014)]%
        {jackson2014policy}
\bibfield{author}{\bibinfo{person}{Steven~J Jackson}, \bibinfo{person}{Tarleton Gillespie}, {and} \bibinfo{person}{Sandy Payette}.} \bibinfo{year}{2014}\natexlab{}.
\newblock \showarticletitle{The policy knot: Re-integrating policy, practice and design in CSCW studies of social computing}. In \bibinfo{booktitle}{\emph{Proceedings of the 17th ACM conference on Computer supported cooperative work \& social computing}}. \bibinfo{pages}{588--602}.
\newblock


\bibitem[Jiang and Lou(2017)]%
        {jiang2017collaborative}
\bibfield{author}{\bibinfo{person}{Chenhan Jiang} {and} \bibinfo{person}{Yongqi Lou}.} \bibinfo{year}{2017}\natexlab{}.
\newblock \showarticletitle{Collaborative Service for Cross-Geographical Design Context: The Case of Sino-Italian Digital Platform}. In \bibinfo{booktitle}{\emph{Cross-Cultural Design: 9th International Conference, CCD 2017, Held as Part of HCI International 2017, Vancouver, BC, Canada, July 9-14, 2017, Proceedings 9}}. Springer, \bibinfo{pages}{345--355}.
\newblock


\bibitem[Kang and Rzeszotarski(2024)]%
        {kang2024challenges}
\bibfield{author}{\bibinfo{person}{Daye Kang} {and} \bibinfo{person}{Jeffrey~M Rzeszotarski}.} \bibinfo{year}{2024}\natexlab{}.
\newblock \showarticletitle{Challenges and Opportunities for Tool Adoption in Industrial UX Research Collaborations}.
\newblock \bibinfo{journal}{\emph{Proceedings of the ACM on Human-Computer Interaction}} \bibinfo{volume}{8}, \bibinfo{number}{CSCW2} (\bibinfo{year}{2024}), \bibinfo{pages}{1--27}.
\newblock


\bibitem[Kashfi et~al\mbox{.}(2017)]%
        {kashfi2017integrating}
\bibfield{author}{\bibinfo{person}{Pariya Kashfi}, \bibinfo{person}{Agneta Nilsson}, {and} \bibinfo{person}{Robert Feldt}.} \bibinfo{year}{2017}\natexlab{}.
\newblock \showarticletitle{Integrating User eXperience practices into software development processes: implications of the UX characteristics}.
\newblock \bibinfo{journal}{\emph{PeerJ Computer Science}}  \bibinfo{volume}{3} (\bibinfo{year}{2017}), \bibinfo{pages}{e130}.
\newblock


\bibitem[Khurana et~al\mbox{.}(2015)]%
        {khurana2015protective}
\bibfield{author}{\bibinfo{person}{Atika Khurana}, \bibinfo{person}{Amy Bleakley}, \bibinfo{person}{Amy~B Jordan}, {and} \bibinfo{person}{Daniel Romer}.} \bibinfo{year}{2015}\natexlab{}.
\newblock \showarticletitle{The protective effects of parental monitoring and internet restriction on adolescents’ risk of online harassment}.
\newblock \bibinfo{journal}{\emph{Journal of youth and Adolescence}}  \bibinfo{volume}{44} (\bibinfo{year}{2015}), \bibinfo{pages}{1039--1047}.
\newblock


\bibitem[Lazar et~al\mbox{.}(2017)]%
        {Lazar_2017}
\bibfield{author}{\bibinfo{person}{Jonathan Lazar}, \bibinfo{person}{Jinjuan~Heidi Feng}, {and} \bibinfo{person}{Harry Hochheiser}.} \bibinfo{year}{2017}\natexlab{}.
\newblock \bibinfo{booktitle}{\emph{Research Methods in Human-Computer Interaction} (\bibinfo{edition}{2} ed.)}.
\newblock \bibinfo{publisher}{Morgan Kaufmann}, \bibinfo{address}{Cambridge, MA}.
\newblock
\showISBNx{978-0-12-805390-4}
\urldef\tempurl%
\url{https://www.safaribooksonline.com/library/view/research-methods-in/9780128093436/}
\showURL{%
\tempurl}


\bibitem[Lee et~al\mbox{.}(2009)]%
        {lee2009}
\bibfield{author}{\bibinfo{person}{Sang-Yong~Tom Lee}, \bibinfo{person}{Hee-Woong Kim}, {and} \bibinfo{person}{Sumeet Gupta}.} \bibinfo{year}{2009}\natexlab{}.
\newblock \showarticletitle{Measuring open source software success}.
\newblock \bibinfo{journal}{\emph{Omega}} \bibinfo{volume}{37}, \bibinfo{number}{2} (\bibinfo{year}{2009}), \bibinfo{pages}{426--438}.
\newblock
\showISSN{0305-0483}
\urldef\tempurl%
\url{https://doi.org/10.1016/j.omega.2007.05.005}
\showDOI{\tempurl}


\bibitem[Lowman(2016)]%
        {Lowman2016}
\bibfield{author}{\bibinfo{person}{RL Lowman}.} \bibinfo{year}{2016}\natexlab{}.
\newblock \showarticletitle{An Introduction to Consulting Psychology: Working with Individuals}.
\newblock \bibinfo{journal}{\emph{Groups, and Organizations}} (\bibinfo{year}{2016}), \bibinfo{pages}{11--70}.
\newblock


\bibitem[Lukáš(2024)]%
        {lukas2024}
\bibfield{author}{\bibinfo{person}{Filip Lukáš}.} \bibinfo{year}{2024}\natexlab{}.
\newblock \bibinfo{booktitle}{\emph{Internet Governance Revamped? Global Digital Compact and NETmundial+10}}.
\newblock CENTR.
\newblock
\urldef\tempurl%
\url{https://www.centr.org/news/blog/internet-governance-revamped-global-digital-compact-and-netmundial-10.html}
\showURL{%
Retrieved May 30, 2024 from \tempurl}


\bibitem[Mao et~al\mbox{.}(2019)]%
        {Mao2019}
\bibfield{author}{\bibinfo{person}{Yaoli Mao}, \bibinfo{person}{Dakuo Wang}, \bibinfo{person}{Michael Muller}, \bibinfo{person}{Kush~R. Varshney}, \bibinfo{person}{Ioana Baldini}, \bibinfo{person}{Casey Dugan}, {and} \bibinfo{person}{Aleksandra Mojsilovi\'{c}}.} \bibinfo{year}{2019}\natexlab{}.
\newblock \showarticletitle{How Data ScientistsWork Together With Domain Experts in Scientific Collaborations: To Find The Right Answer Or To Ask The Right Question?}
\newblock \bibinfo{journal}{\emph{Proc. ACM Hum.-Comput. Interact.}} \bibinfo{volume}{3}, \bibinfo{number}{GROUP}, Article \bibinfo{articleno}{237} (\bibinfo{date}{Dec.} \bibinfo{year}{2019}), \bibinfo{numpages}{23}~pages.
\newblock
\urldef\tempurl%
\url{https://doi.org/10.1145/3361118}
\showDOI{\tempurl}


\bibitem[Masaki et~al\mbox{.}(2020)]%
        {masaki2020exploring}
\bibfield{author}{\bibinfo{person}{Hiroaki Masaki}, \bibinfo{person}{Kengo Shibata}, \bibinfo{person}{Shui Hoshino}, \bibinfo{person}{Takahiro Ishihama}, \bibinfo{person}{Nagayuki Saito}, {and} \bibinfo{person}{Koji Yatani}.} \bibinfo{year}{2020}\natexlab{}.
\newblock \showarticletitle{Exploring nudge designs to help adolescent sns users avoid privacy and safety threats}. In \bibinfo{booktitle}{\emph{Proceedings of the 2020 CHI Conference on Human Factors in Computing Systems}}. \bibinfo{pages}{1--11}.
\newblock


\bibitem[Munson et~al\mbox{.}(2013)]%
        {munson2013sociotechnical}
\bibfield{author}{\bibinfo{person}{Sean~A Munson}, \bibinfo{person}{Hasan Cavusoglu}, \bibinfo{person}{Larry Frisch}, {and} \bibinfo{person}{Sidney Fels}.} \bibinfo{year}{2013}\natexlab{}.
\newblock \showarticletitle{Sociotechnical challenges and progress in using social media for health}.
\newblock \bibinfo{journal}{\emph{Journal of medical Internet research}} \bibinfo{volume}{15}, \bibinfo{number}{10} (\bibinfo{year}{2013}), \bibinfo{pages}{e226}.
\newblock


\bibitem[NETmundial(2014)]%
        {NETmundial2014}
\bibfield{author}{\bibinfo{person}{NETmundial}.} \bibinfo{year}{2014}\natexlab{}.
\newblock \bibinfo{booktitle}{\emph{NETmundial+10 Multistakeholder Statement}}.
\newblock NETmundial.
\newblock
\urldef\tempurl%
\url{https://netmundial.br/2014/wp-content/uploads/2014/04/NETmundial-Multistakeholder-Document.pdf}
\showURL{%
Retrieved May 30, 2024 from \tempurl}


\bibitem[NETmundial(2024)]%
        {NETmundial2024}
\bibfield{author}{\bibinfo{person}{NETmundial}.} \bibinfo{year}{2024}\natexlab{}.
\newblock \bibinfo{booktitle}{\emph{NETmundial+10 Multistakeholder Statement}}.
\newblock NETmundial.
\newblock
\urldef\tempurl%
\url{https://netmundial.br/pdf/NETmundial10-MultistakeholderStatement-2024.pdf}
\showURL{%
Retrieved May 30, 2024 from \tempurl}


\bibitem[Nomura et~al\mbox{.}(2008)]%
        {Nomura2008}
\bibfield{author}{\bibinfo{person}{Saeko Nomura}, \bibinfo{person}{Jeremy Birnholtz}, \bibinfo{person}{Oya Rieger}, \bibinfo{person}{Gilly Leshed}, \bibinfo{person}{Deborah Trumbull}, {and} \bibinfo{person}{Geri Gay}.} \bibinfo{year}{2008}\natexlab{}.
\newblock \showarticletitle{Cutting into collaboration: understanding coordination in distributed and interdisciplinary medical research}. In \bibinfo{booktitle}{\emph{Proceedings of the 2008 ACM Conference on Computer Supported Cooperative Work}} (San Diego, CA, USA) \emph{(\bibinfo{series}{CSCW '08})}. \bibinfo{publisher}{Association for Computing Machinery}, \bibinfo{address}{New York, NY, USA}, \bibinfo{pages}{427–436}.
\newblock
\showISBNx{9781605580074}
\urldef\tempurl%
\url{https://doi.org/10.1145/1460563.1460632}
\showDOI{\tempurl}


\bibitem[of~the Surgeon General~(OSG)(2023)]%
        {OSG2023_advisory}
\bibfield{author}{\bibinfo{person}{Office of~the Surgeon General~(OSG)}.} \bibinfo{year}{2023}\natexlab{}.
\newblock \bibinfo{title}{Social Media and Youth Mental Health: The U.S. Surgeon General’s Advisory}.
\newblock
\newblock
\urldef\tempurl%
\url{https://www.hhs.gov/sites/default/files/sg-youth-mental-health-social-media-advisory.pdf}
\showURL{%
\tempurl}


\bibitem[Olson and Olson(2000)]%
        {Olson2000}
\bibfield{author}{\bibinfo{person}{Gary~M. Olson} {and} \bibinfo{person}{Judith~S. Olson}.} \bibinfo{year}{2000}\natexlab{}.
\newblock \showarticletitle{Distance matters}.
\newblock \bibinfo{journal}{\emph{Hum.-Comput. Interact.}} \bibinfo{volume}{15}, \bibinfo{number}{2} (\bibinfo{date}{Sept.} \bibinfo{year}{2000}), \bibinfo{pages}{139–178}.
\newblock
\showISSN{0737-0024}
\urldef\tempurl%
\url{https://doi.org/10.1207/S15327051HCI1523_4}
\showDOI{\tempurl}


\bibitem[Olson et~al\mbox{.}(2009)]%
        {Olson2009}
\bibfield{author}{\bibinfo{person}{Gary~M Olson}, \bibinfo{person}{Judith~S Olson}, {and} \bibinfo{person}{Gina Venolia}.} \bibinfo{year}{2009}\natexlab{}.
\newblock \showarticletitle{What still matters about distance}.
\newblock \bibinfo{journal}{\emph{Proceedings of HCIC}} (\bibinfo{year}{2009}).
\newblock
\urldef\tempurl%
\url{https://www.microsoft.com/en-us/research/wp-content/uploads/2016/02/Olson9370.pdf}
\showURL{%
\tempurl}


\bibitem[O'Mahony and Karp(2022)]%
        {omahony2022proprietary}
\bibfield{author}{\bibinfo{person}{Siobhan O'Mahony} {and} \bibinfo{person}{Rebecca Karp}.} \bibinfo{year}{2022}\natexlab{}.
\newblock \showarticletitle{From proprietary to collective governance: How do platform participation strategies evolve?}
\newblock \bibinfo{journal}{\emph{Strategic Management Journal}} \bibinfo{volume}{43}, \bibinfo{number}{3} (\bibinfo{year}{2022}), \bibinfo{pages}{530--562}.
\newblock
\urldef\tempurl%
\url{https://doi.org/10.1002/smj.3150}
\showDOI{\tempurl}


\bibitem[Park et~al\mbox{.}(2024)]%
        {park2024personally}
\bibfield{author}{\bibinfo{person}{Jinkyung Park}, \bibinfo{person}{Joshua Gracie}, \bibinfo{person}{Ashwaq Alsoubai}, \bibinfo{person}{Afsaneh Razi}, {and} \bibinfo{person}{Pamela~J Wisniewski}.} \bibinfo{year}{2024}\natexlab{}.
\newblock \showarticletitle{Personally Targeted Risk vs. Humor: How Online Risk Perceptions of Youth vs. Third-Party Annotators Differ based on Privately Shared Media on Instagram}. In \bibinfo{booktitle}{\emph{IDC'24: Proceedings of the 23rd Annual ACM Interaction Design and Children Conference}}. Association for Computing Machinery, \bibinfo{pages}{1--13}.
\newblock


\bibitem[Park et~al\mbox{.}(2023)]%
        {park2023towards}
\bibfield{author}{\bibinfo{person}{Jinkyung Park}, \bibinfo{person}{Joshua Gracie}, \bibinfo{person}{Ashwaq Alsoubai}, \bibinfo{person}{Gianluca Stringhini}, \bibinfo{person}{Vivek Singh}, {and} \bibinfo{person}{Pamela Wisniewski}.} \bibinfo{year}{2023}\natexlab{}.
\newblock \showarticletitle{Towards automated detection of risky images shared by youth on social media}. In \bibinfo{booktitle}{\emph{Companion Proceedings of the ACM Web Conference 2023}}. \bibinfo{pages}{1348--1357}.
\newblock


\bibitem[Qadir et~al\mbox{.}(2024)]%
        {qadir2024towards}
\bibfield{author}{\bibinfo{person}{Sarvech Qadir}, \bibinfo{person}{Andy Niser}, \bibinfo{person}{Xavier Caddle}, \bibinfo{person}{Ashwaq Alsoubai}, \bibinfo{person}{Jinkyung~Katie Park}, {and} \bibinfo{person}{Pamela~J Wisniewski}.} \bibinfo{year}{2024}\natexlab{}.
\newblock \showarticletitle{Towards a Safer Digital Future: Exploring Secondary Stakeholder Perspectives on Creating a Sustainable Youth Online Safety Community}.
\newblock  (\bibinfo{year}{2024}).
\newblock


\bibitem[Razi et~al\mbox{.}(2023)]%
        {razi2023sliding}
\bibfield{author}{\bibinfo{person}{Afsaneh Razi}, \bibinfo{person}{Ashwaq AlSoubai}, \bibinfo{person}{Seunghyun Kim}, \bibinfo{person}{Shiza Ali}, \bibinfo{person}{Gianluca Stringhini}, \bibinfo{person}{Munmun De~Choudhury}, {and} \bibinfo{person}{Pamela~J Wisniewski}.} \bibinfo{year}{2023}\natexlab{}.
\newblock \showarticletitle{Sliding into my DMs: Detecting uncomfortable or unsafe sexual risk experiences within Instagram direct messages grounded in the perspective of youth}.
\newblock \bibinfo{journal}{\emph{Proceedings of the ACM on Human-Computer Interaction}} \bibinfo{volume}{7}, \bibinfo{number}{CSCW1} (\bibinfo{year}{2023}), \bibinfo{pages}{1--29}.
\newblock


\bibitem[Razi et~al\mbox{.}(2022)]%
        {razi2022instagram}
\bibfield{author}{\bibinfo{person}{Afsaneh Razi}, \bibinfo{person}{Ashwaq AlSoubai}, \bibinfo{person}{Seunghyun Kim}, \bibinfo{person}{Nurun Naher}, \bibinfo{person}{Shiza Ali}, \bibinfo{person}{Gianluca Stringhini}, \bibinfo{person}{Munmun De~Choudhury}, {and} \bibinfo{person}{Pamela~J Wisniewski}.} \bibinfo{year}{2022}\natexlab{}.
\newblock \showarticletitle{Instagram data donation: a case study on collecting ecologically valid social media data for the purpose of adolescent online risk detection}. In \bibinfo{booktitle}{\emph{CHI Conference on Human Factors in Computing Systems Extended Abstracts}}. \bibinfo{pages}{1--9}.
\newblock


\bibitem[Razi et~al\mbox{.}(2020)]%
        {Razi_lets_2020}
\bibfield{author}{\bibinfo{person}{Afsaneh Razi}, \bibinfo{person}{Karla Badillo-Urquiola}, {and} \bibinfo{person}{Pamela~J. Wisniewski}.} \bibinfo{year}{2020}\natexlab{}.
\newblock \showarticletitle{Let's {Talk} about {Sext}: {How} {Adolescents} {Seek} {Support} and {Advice} about {Their} {Online} {Sexual} {Experiences}}. In \bibinfo{booktitle}{\emph{Proceedings of the 2020 {CHI} {Conference} on {Human} {Factors} in {Computing} {Systems}}} \emph{(\bibinfo{series}{{CHI} '20})}. \bibinfo{publisher}{Association for Computing Machinery}, \bibinfo{address}{Honolulu, HI, USA}, \bibinfo{pages}{1--13}.
\newblock
\showISBNx{978-1-4503-6708-0}
\urldef\tempurl%
\url{https://doi.org/10.1145/3313831.3376400}
\showDOI{\tempurl}


\bibitem[R{\"u}ller et~al\mbox{.}(2021)]%
        {ruller2021technology}
\bibfield{author}{\bibinfo{person}{Sarah R{\"u}ller}, \bibinfo{person}{Konstantin Aal}, \bibinfo{person}{Simon Holdermann}, \bibinfo{person}{Peter Tolmie}, \bibinfo{person}{Andrea Hartmann}, \bibinfo{person}{Markus Rohde}, \bibinfo{person}{Martin Zillinger}, {and} \bibinfo{person}{Volker Wulf}.} \bibinfo{year}{2021}\natexlab{}.
\newblock \showarticletitle{‘Technology is everywhere, we have the opportunity to learn it in the valley’: the appropriation of a socio-technical enabling infrastructure in the Moroccan high atlas}.
\newblock \bibinfo{journal}{\emph{Computer Supported Cooperative Work (CSCW)}} (\bibinfo{year}{2021}), \bibinfo{pages}{1--40}.
\newblock


\bibitem[Rutkowski et~al\mbox{.}(2021)]%
        {rutkowski2021family}
\bibfield{author}{\bibinfo{person}{Tara~L Rutkowski}, \bibinfo{person}{Heidi Hartikainen}, \bibinfo{person}{Kirsten~E Richards}, {and} \bibinfo{person}{Pamela~J Wisniewski}.} \bibinfo{year}{2021}\natexlab{}.
\newblock \showarticletitle{Family communication: examining the differing perceptions of parents and teens regarding online safety communication}.
\newblock \bibinfo{journal}{\emph{Proceedings of the ACM on Human-Computer Interaction}} \bibinfo{volume}{5}, \bibinfo{number}{CSCW2} (\bibinfo{year}{2021}), \bibinfo{pages}{1--23}.
\newblock


\bibitem[Ryan and Deci(2000)]%
        {ryan2000self}
\bibfield{author}{\bibinfo{person}{Richard~M Ryan} {and} \bibinfo{person}{Edward~L Deci}.} \bibinfo{year}{2000}\natexlab{}.
\newblock \showarticletitle{Self-determination theory and the facilitation of intrinsic motivation, social development, and well-being.}
\newblock \bibinfo{journal}{\emph{American psychologist}} \bibinfo{volume}{55}, \bibinfo{number}{1} (\bibinfo{year}{2000}), \bibinfo{pages}{68}.
\newblock


\bibitem[Ryan and Deci(2020)]%
        {ryan2020intrinsic}
\bibfield{author}{\bibinfo{person}{Richard~M Ryan} {and} \bibinfo{person}{Edward~L Deci}.} \bibinfo{year}{2020}\natexlab{}.
\newblock \showarticletitle{Intrinsic and extrinsic motivation from a self-determination theory perspective: Definitions, theory, practices, and future directions}.
\newblock \bibinfo{journal}{\emph{Contemporary educational psychology}}  \bibinfo{volume}{61} (\bibinfo{year}{2020}), \bibinfo{pages}{101860}.
\newblock


\bibitem[Saha et~al\mbox{.}(2022)]%
        {saha2022towards}
\bibfield{author}{\bibinfo{person}{Manika Saha}, \bibinfo{person}{Delvin Varghese}, \bibinfo{person}{Tom Bartindale}, \bibinfo{person}{Shakuntala~Haraksingh Thilsted}, \bibinfo{person}{Syed~Ishtiaque Ahmed}, {and} \bibinfo{person}{Patrick Olivier}.} \bibinfo{year}{2022}\natexlab{}.
\newblock \showarticletitle{Towards sustainable ICTD in Bangladesh: understanding the program and policy landscape and its implications for CSCW and HCI}.
\newblock \bibinfo{journal}{\emph{Proceedings of the ACM on Human-Computer Interaction}} \bibinfo{volume}{6}, \bibinfo{number}{CSCW1} (\bibinfo{year}{2022}), \bibinfo{pages}{1--31}.
\newblock


\bibitem[Schiano and Burg(2017)]%
        {schiano2017parental}
\bibfield{author}{\bibinfo{person}{Diane~J Schiano} {and} \bibinfo{person}{Christine Burg}.} \bibinfo{year}{2017}\natexlab{}.
\newblock \showarticletitle{Parental controls: Oxymoron and design opportunity}. In \bibinfo{booktitle}{\emph{HCI International 2017--Posters' Extended Abstracts: 19th International Conference, HCI International 2017, Vancouver, BC, Canada, July 9--14, 2017, Proceedings, Part II 19}}. Springer, \bibinfo{pages}{645--652}.
\newblock


\bibitem[Schoenebeck et~al\mbox{.}(2021)]%
        {Schoenebeck2021}
\bibfield{author}{\bibinfo{person}{Sarita Schoenebeck}, \bibinfo{person}{Carol~F. Scott}, \bibinfo{person}{Emma~Grace Hurley}, \bibinfo{person}{Tammy Chang}, {and} \bibinfo{person}{Ellen Selkie}.} \bibinfo{year}{2021}\natexlab{}.
\newblock \showarticletitle{Youth Trust in Social Media Companies and Expectations of Justice: Accountability and Repair After Online Harassment}.
\newblock \bibinfo{journal}{\emph{Proc. ACM Hum.-Comput. Interact.}} \bibinfo{volume}{5}, \bibinfo{number}{CSCW1}, Article \bibinfo{articleno}{2} (\bibinfo{date}{apr} \bibinfo{year}{2021}), \bibinfo{numpages}{18}~pages.
\newblock
\urldef\tempurl%
\url{https://doi.org/10.1145/3449076}
\showDOI{\tempurl}


\bibitem[Schweik and English(2012)]%
        {schweik2012}
\bibfield{author}{\bibinfo{person}{Charles~M Schweik} {and} \bibinfo{person}{Robert~C English}.} \bibinfo{year}{2012}\natexlab{}.
\newblock \bibinfo{booktitle}{\emph{Internet success: a study of open-source software commons}}.
\newblock \bibinfo{publisher}{MIT Press}.
\newblock


\bibitem[Shibuya and Tamai(2009)]%
        {shibuya2009}
\bibfield{author}{\bibinfo{person}{Bianca Shibuya} {and} \bibinfo{person}{Tetsuo Tamai}.} \bibinfo{year}{2009}\natexlab{}.
\newblock \showarticletitle{Understanding the process of participating in open source communities}. In \bibinfo{booktitle}{\emph{2009 ICSE Workshop on Emerging Trends in Free/Libre/Open Source Software Research and Development}}. \bibinfo{pages}{1--6}.
\newblock
\urldef\tempurl%
\url{https://doi.org/10.1109/FLOSS.2009.5071352}
\showDOI{\tempurl}


\bibitem[Shneiderman et~al\mbox{.}(2016)]%
        {Shneiderman2016}
\bibfield{author}{\bibinfo{person}{Ben Shneiderman}, \bibinfo{person}{Catherine Plaisant}, \bibinfo{person}{Maxine Cohen}, \bibinfo{person}{Steven Jacobs}, \bibinfo{person}{Niklas Elmqvist}, {and} \bibinfo{person}{Nicholas Diakopoulos}.} \bibinfo{year}{2016}\natexlab{}.
\newblock \bibinfo{booktitle}{\emph{Designing the User Interface: Strategies for Effective Human-Computer Interaction} (\bibinfo{edition}{6th} ed.)}.
\newblock \bibinfo{publisher}{Pearson}.
\newblock
\showISBNx{013438038X}


\bibitem[Simon(2005)]%
        {simon2005}
\bibfield{author}{\bibinfo{person}{K.~D. Simon}.} \bibinfo{year}{2005}\natexlab{}.
\newblock \showarticletitle{The value of open standards and open-source software in government environments}.
\newblock \bibinfo{journal}{\emph{IBM Systems Journal}} \bibinfo{volume}{44}, \bibinfo{number}{2} (\bibinfo{year}{2005}), \bibinfo{pages}{227--238}.
\newblock
\showISBNx{00188670}
\urldef\tempurl%
\url{https://www.proquest.com/scholarly-journals/value-open-standards-source-software-government/docview/222420968/se-2}
\showURL{%
\tempurl}
\newblock
\shownote{Copyright - Copyright International Business Machines Corporation 2005; Document feature - graphs; references; Last updated - 2022-10-20; CODEN - IBMSA7}.


\bibitem[S{\o}rensen and Torfing(2009)]%
        {sorensen2009making}
\bibfield{author}{\bibinfo{person}{Eva S{\o}rensen} {and} \bibinfo{person}{Jacob Torfing}.} \bibinfo{year}{2009}\natexlab{}.
\newblock \showarticletitle{Making governance networks effective and democratic through metagovernance}.
\newblock \bibinfo{journal}{\emph{Public administration}} \bibinfo{volume}{87}, \bibinfo{number}{2} (\bibinfo{year}{2009}), \bibinfo{pages}{234--258}.
\newblock


\bibitem[Svirina et~al\mbox{.}(2016)]%
        {svirina2016implementing}
\bibfield{author}{\bibinfo{person}{Anna Svirina}, \bibinfo{person}{Alfia Zabbarova}, {and} \bibinfo{person}{Karine Oganisjana}.} \bibinfo{year}{2016}\natexlab{}.
\newblock \showarticletitle{Implementing open innovation concept in social business}.
\newblock \bibinfo{journal}{\emph{Journal of open innovation: Technology, market, and Complexity}} \bibinfo{volume}{2}, \bibinfo{number}{4} (\bibinfo{year}{2016}), \bibinfo{pages}{1--10}.
\newblock


\bibitem[Sweigart et~al\mbox{.}(2025)]%
        {sweigart2025takes}
\bibfield{author}{\bibinfo{person}{Elizabeth~A Sweigart}, \bibinfo{person}{Jinkyung~Katie Park}, {and} \bibinfo{person}{Pamela~J Wisniewski}.} \bibinfo{year}{2025}\natexlab{}.
\newblock \showarticletitle{It Takes a Village: Youth Online Safety Research Highlights Need for Interdisciplinary, Multistakeholder Solutions}.
\newblock \bibinfo{journal}{\emph{Journal of Advances in Information Technology}} \bibinfo{volume}{16}, \bibinfo{number}{1} (\bibinfo{year}{2025}).
\newblock


\bibitem[Tongco(2007)]%
        {Tongco_2007}
\bibfield{author}{\bibinfo{person}{Ma~Dolores~C Tongco}.} \bibinfo{year}{2007}\natexlab{}.
\newblock \showarticletitle{Purposive sampling as a tool for informant selection}.
\newblock \bibinfo{journal}{\emph{Ethnobotany Research and applications}}  \bibinfo{volume}{5} (\bibinfo{year}{2007}), \bibinfo{pages}{147--158}.
\newblock


\bibitem[Truong et~al\mbox{.}(2022)]%
        {rruong2022}
\bibfield{author}{\bibinfo{person}{Kimberly Truong}, \bibinfo{person}{Courtney Miller}, \bibinfo{person}{Bogdan Vasilescu}, {and} \bibinfo{person}{Christian Kästner}.} \bibinfo{year}{2022}\natexlab{}.
\newblock \showarticletitle{The Unsolvable Problem or the Unheard Answer? A Dataset of 24,669 Open-Source Software Conference Talks}. In \bibinfo{booktitle}{\emph{2022 IEEE/ACM 19th International Conference on Mining Software Repositories (MSR)}}. \bibinfo{pages}{348--352}.
\newblock
\urldef\tempurl%
\url{https://doi.org/10.1145/3524842.3528488}
\showDOI{\tempurl}


\bibitem[Vlachokyriakos et~al\mbox{.}(2021)]%
        {vlachokyriakos2021research}
\bibfield{author}{\bibinfo{person}{Vasilis Vlachokyriakos}, \bibinfo{person}{Clara Crivellaro}, \bibinfo{person}{Hara Kouki}, \bibinfo{person}{Christos Giovanopoulos}, {and} \bibinfo{person}{Patrick Olivier}.} \bibinfo{year}{2021}\natexlab{}.
\newblock \showarticletitle{Research with a solidarity clinic: Design implications for CSCW Healthcare Service Design}.
\newblock \bibinfo{journal}{\emph{Computer Supported Cooperative Work (CSCW)}} \bibinfo{volume}{30}, \bibinfo{number}{5} (\bibinfo{year}{2021}), \bibinfo{pages}{757--783}.
\newblock


\bibitem[Wallace et~al\mbox{.}(2013)]%
        {wallace2013}
\bibfield{author}{\bibinfo{person}{Jayne Wallace}, \bibinfo{person}{John McCarthy}, \bibinfo{person}{Peter~C Wright}, {and} \bibinfo{person}{Patrick Olivier}.} \bibinfo{year}{2013}\natexlab{}.
\newblock \showarticletitle{Making design probes work}. In \bibinfo{booktitle}{\emph{Proceedings of the SIGCHI Conference on Human Factors in Computing Systems}}. \bibinfo{pages}{3441--3450}.
\newblock


\bibitem[Wang et~al\mbox{.}(2022)]%
        {Wang2022}
\bibfield{author}{\bibinfo{person}{Dakuo Wang}, \bibinfo{person}{Michael Muller}, \bibinfo{person}{Qian Yang}, \bibinfo{person}{Zijun Wang}, \bibinfo{person}{Ming Tan}, {and} \bibinfo{person}{Stacy Hobson}.} \bibinfo{year}{2022}\natexlab{}.
\newblock \showarticletitle{Organizational Distance Also Matters: How Organizational Distance Among Industrial Research Teams Affect Their Research Productivity}.
\newblock \bibinfo{journal}{\emph{Proc. ACM Hum.-Comput. Interact.}} \bibinfo{volume}{6}, \bibinfo{number}{CSCW2}, Article \bibinfo{articleno}{453} (\bibinfo{date}{Nov.} \bibinfo{year}{2022}), \bibinfo{numpages}{18}~pages.
\newblock
\urldef\tempurl%
\url{https://doi.org/10.1145/3555554}
\showDOI{\tempurl}


\bibitem[Weiss and Bailetti(2015)]%
        {weiss2015}
\bibfield{author}{\bibinfo{person}{Michael Weiss} {and} \bibinfo{person}{Tony Bailetti}.} \bibinfo{year}{2015}\natexlab{}.
\newblock \showarticletitle{Value of open source projects: A case for open source cybersecurity}. In \bibinfo{booktitle}{\emph{2015 IEEE International Conference on Engineering, Technology and Innovation/ International Technology Management Conference (ICE/ITMC)}}. \bibinfo{pages}{1--8}.
\newblock
\urldef\tempurl%
\url{https://doi.org/10.1109/ICE.2015.7438667}
\showDOI{\tempurl}


\bibitem[Wisniewski et~al\mbox{.}(2017)]%
        {wisniewski2017parental}
\bibfield{author}{\bibinfo{person}{Pamela Wisniewski}, \bibinfo{person}{Arup~Kumar Ghosh}, \bibinfo{person}{Heng Xu}, \bibinfo{person}{Mary~Beth Rosson}, {and} \bibinfo{person}{John~M Carroll}.} \bibinfo{year}{2017}\natexlab{}.
\newblock \showarticletitle{Parental control vs. teen self-regulation: Is there a middle ground for mobile online safety?}. In \bibinfo{booktitle}{\emph{Proceedings of the 2017 ACM conference on computer supported cooperative work and social computing}}. \bibinfo{pages}{51--69}.
\newblock


\bibitem[Wisniewski et~al\mbox{.}(2025)]%
        {wisniewski2025moving}
\bibfield{author}{\bibinfo{person}{Pamela Wisniewski}, \bibinfo{person}{Jinkyung Park}, \bibinfo{person}{Karla Badillo-Urquiola}, \bibinfo{person}{Joy Gabrielli}, \bibinfo{person}{Jennifer~L Doty}, {and} \bibinfo{person}{Heidi Hartikainen}.} \bibinfo{year}{2025}\natexlab{}.
\newblock \showarticletitle{Moving Beyond Fear and Restriction to Promoting Adolescent Resilience and Intentional Technology Use}.
\newblock In \bibinfo{booktitle}{\emph{Handbook of Children and Screens}}. \bibinfo{publisher}{Springer, Cham}, \bibinfo{pages}{403--410}.
\newblock


\bibitem[Wisniewski et~al\mbox{.}(2022)]%
        {wisniewski2022privacy}
\bibfield{author}{\bibinfo{person}{Pamela~J Wisniewski}, \bibinfo{person}{Jessica Vitak}, {and} \bibinfo{person}{Heidi Hartikainen}.} \bibinfo{year}{2022}\natexlab{}.
\newblock \showarticletitle{Privacy in adolescence}.
\newblock In \bibinfo{booktitle}{\emph{Modern socio-technical perspectives on privacy}}. \bibinfo{publisher}{Springer International Publishing Cham}, \bibinfo{pages}{315--336}.
\newblock


\bibitem[Zhang et~al\mbox{.}(2020)]%
        {Zhang2020}
\bibfield{author}{\bibinfo{person}{Amy~X. Zhang}, \bibinfo{person}{Michael Muller}, {and} \bibinfo{person}{Dakuo Wang}.} \bibinfo{year}{2020}\natexlab{}.
\newblock \showarticletitle{How do Data Science Workers Collaborate? Roles, Workflows, and Tools}.
\newblock \bibinfo{journal}{\emph{Proc. ACM Hum.-Comput. Interact.}} \bibinfo{volume}{4}, \bibinfo{number}{CSCW1}, Article \bibinfo{articleno}{22} (\bibinfo{date}{May} \bibinfo{year}{2020}), \bibinfo{numpages}{23}~pages.
\newblock
\urldef\tempurl%
\url{https://doi.org/10.1145/3392826}
\showDOI{\tempurl}


\end{thebibliography}

\appendix
\pagebreak \raggedbottom
\section{SESSION FLOW AND SEMI-STRUCTURED INTERVIEW QUESTIONS}\label{sec.study3-script}\label{sec.interviewquestions-study3}
Below you will find the interview questions that were used for our semi-scripted interviews. The interviews were conducted as part of a project funded by [anonymized for review]. 

\begin{center}\underline{\textbf{Introduction and Consent}}\end{center}
\begin{enumerate}
    \item Introduce yourself
    \item Ask permission to use for research purposes
    \item Ask permission to record
\end{enumerate}

\begin{center}\underline{\textbf{Section 1: Dashboard Design Probe}}\end{center}

Video plays (2 minutes 24 seconds)

\textit{\textbf{Interviewer}}:
\begin{itemize}
\item What are your overall thoughts about the dashboard that was presented? 

\item Do you think there will be any barriers that would not let community members contribute to this dashboard? 

\item Who are other stakeholders or populations that could benefit from this dashboard other than youth?

\item Do you think there will be any barriers that would not let youth benefit from this dashboard?

\item What dashboard features do you think are useful for youth or other community members? 

\item What features could be added to the dashboard to enhance the experience of users? 

\item Should access to the dashboard be free?
\end{itemize}

\begin{center}\underline{\textbf{Section 2: Community Walkthrough}}\end{center}

\textbf{Task 1: Member registration and profile creation. Walk through the steps of creating a contributor profile.}

\begin{itemize}
\item How could we improve the profile creation process?

\item What information would be important for the community to capture upon profile creation?

\item What details would you want to showcase on your profile?

\item What types of information would you not be willing to share?
\end{itemize}

\textbf{Task 2: Member search. Find a community member that you might be interested in contacting to work with on a future project.}

\begin{itemize}
    \item For what purpose would you see yourself looking for people to expand your network? What kinds of projects?

    \item Based on the current member profiles, can you identify someone who would look interesting for you to work with and why?

    \item What search fields do you think would be important to search by to achieve this?

    \item If another member wanted to partner with you through the community, what would be the best way for them to go about doing that even if that method is not currently implemented?
\end{itemize}

\textbf{Task 3: Projects. Use the projects section to propose a project to the community.}
\begin{itemize}
    \item What types of projects would you foresee wanting to collaborate on with community members?

    \item What details would be useful for capture from users to provide for such a project?

    \item What barriers would prevent users from proposing projects?

    \item Are there any regulations or rules that should be added to the community to ensure that no one can steal someone's ideas or projects?
\end{itemize}

\textbf{Task 4: Post on the forum}
\begin{itemize}
    \item What types of topics would you like to see on a forum like this?

    \item What topics would not be interesting to you?

    \item What would encourage you to engage in community discussions?

    \item What would stop you from engaging in community discussions?
\end{itemize}

\begin{center}\underline{\textbf{Section Closing Questions}}\end{center}
\begin{itemize}
    \item What tangible steps can we take to start building this community?
    
    \item What would it take to get you to be one of the contributors to the community?

    \item Are there any existing platforms that you believe would work better than the current one for Open Source community building?

    \item Is there a benefit to doing this on our own versus using LinkedIn or another platform that already exists?

    \item What did you not like about the community? What could be done to improve this area?

    \item If you could add anything to the community, what would it be? Why would you make this modification?

    \item What was missing or disappointing in your experience with the community?

    \item What feature did you value the most and why?

    \item Is there any functionality that should only be available to registered users?

    \item Should registration to this community be moderated or require admin verification before access is allowed?
\end{itemize}

\begin{center}\underline{\textbf{End of Session}}\end{center}
\pagebreak\section{Participant Characteristics}\label{sec.participants-study3}

\newcommand{\STAB}[1]{\begin{tabular}{@{}c@{}}#1\end{tabular}} 
\newcommand{\PreserveBackslash}[1]{\let\temp=\\#1\let\\=\temp}
\begin{table}[htp]
\caption{Participant characteristics}
\label{tab:p-table-study3}
\centering

\begin{tabular}{lll}
\toprule
\textbf{Participant ID} & \textbf{Role}  & \textbf{Job Responsibilities}  \\ \midrule
1                       & Researcher          & Assistant professor       \\ \hline
2                       & Service Provider       & Case manager                    \\ \hline
3                       & Researcher       & Assistant professor           \\ \hline
4                       & Service Provider          & Therapist and Business owner               \\ \hline
5                       & Industry              & Business owner                    \\ \hline
6                       & Service Provider              & Youth advocate                      \\ \hline
7                       & Researcher              & Post-Doctoral Researcher       \\ \hline
8                       & Industry              & Educational software consulting                 \\ \hline
9                       & Industry              & Content Moderation                 \\ \hline
10                      & Service Provider              & Youth advocate                \\ \hline
11                      & Industry       & Business owner               \\ \hline
12                      & Service Provider       & Retired law enforcement             \\ \hline
13                      & Service Provider       & Education administrator         \\ \hline
14                      & Researcher              & Director of youth research firm                        \\ \hline
15                      & Researcher          & Associate professor                  \\ \hline
16                      & Service Provider       & Teacher                     \\ \hline
17                      & Researcher          & Associate Professor                    \\ \hline
18                      & Researcher          & Post-Doctoral Researcher                      \\ \hline
19                      & Researcher              & Associate professor                       \\ \hline
20                      & Researcher          & Associate professor               \\ \hline
21                      & Researcher          & Assistant professor                   \\ \hline
22                      & Industry         & Technical Leader                    \\ \hline
23                      & Service Provider          & Curriculum building             \\ \hline
24                      & Service Provider         & Teacher             \\ \hline
25                      & Industry          & Application development                 \\ \hline
26                      & Service Provider        & Teacher          \\ \hline
27                      & Industry       & Application development           \\ \hline
28                      & Industry       & Business owner                  \\ \hline
29                      & Industry       & Business owner                   \\ \hline
30                      & Service Provider       & Education administrator                 \\ \hline
31                      & Industry       & Business owner                  \\ \hline
32                      & Industry       & Engineer         \\ \hline
33                      & Industry        & Business owner                   \\  \bottomrule
\end{tabular}

\end{table}
\pagebreak\section{Community Design Probe Features and Access Matrix}
\begin{table}[ht]
    \centering
\caption{Initial \acrshort{mosafelyacr} Capability-Support Pairing}
\label{tab:capabilitysupport}
    \begin{tabular}{ll}
        \toprule
         \textbf{Feature}& \textbf{Supports}\\
         \midrule
         News& Disseminating scientifically backed information and updates\\ \hline
         Posts& User interaction and engagement\\ \hline
         Member Search& User interaction\\ \hline
         Projects& Informed design and development of youth online safety solutions\\ \hline
         Workshops& Disseminating scientifically backed information\\
         \bottomrule
    \end{tabular}    
\end{table}

\begin{table}[ht]
    \centering
\caption{\acrshort{mosafelyacr} Feature Set}
\label{tab:featureset}
    \begin{tabular}{lcccc}
    \toprule
         \textbf{Feature}&  \textbf{Guest}&  \textbf{Member}&  \textbf{Contributor}& \textbf{Administrator}\\
         \midrule
         \textbf{Create news}&  &  &  & \checkmark\\ \hline
         \textbf{Browse news}&  \checkmark&  \checkmark&  \checkmark& \checkmark\\ \hline
         \textbf{Comment on news}&  &  \checkmark&  \checkmark& \checkmark\\ \hline
         \textbf{Create posts}&  &  \checkmark&  \checkmark& \checkmark\\ \hline
         \textbf{Browse posts}&  \checkmark&  \checkmark&  \checkmark& \checkmark\\ \hline
 \textbf{Comment on posts}& & \checkmark& \checkmark&\checkmark\\ \hline
 \textbf{Search for members}& & \checkmark& \checkmark&\checkmark\\ \hline
\textbf{Message members}& & \checkmark& \checkmark&\checkmark\\ \hline
 \textbf{Browse projects}& & \checkmark& \checkmark&\checkmark\\ \hline
 \textbf{Propose projects}& & \checkmark& \checkmark&\checkmark\\ \hline
 \textbf{Comment on projects}& & \checkmark& \checkmark&\checkmark\\ \hline 
 \textbf{Access project code repository}& & & \checkmark&\checkmark\\ \hline
 \textbf{Create workshops}& & & &\checkmark\\ \hline
 \textbf{Browse workshops}& \checkmark& \checkmark& \checkmark&\checkmark\\
 \bottomrule
    \end{tabular}    
\end{table}

\end{document}